\documentclass[a4paper,10pt]{article}
\usepackage{enumerate}
\usepackage{amsmath}
\usepackage{amsfonts}
\usepackage{amssymb}
\usepackage{amsthm}
\usepackage{graphicx}
\usepackage{geometry}
\usepackage{float}
\usepackage{graphics}
\usepackage{float}
\usepackage{times}
\usepackage{color}
\usepackage{tikz}
\usetikzlibrary{graphs,graphs.standard}
\usepackage{fontenc}
\usepackage[english]{babel}
\usepackage{fontenc}
\usepackage[english]{babel}
\theoremstyle{definition}
\newtheorem{theorem}{Theorem}
\newtheorem{definition}{Definition}
\newtheorem{axiom}{Axiom}

\newtheorem{remark}{Remark}
\newtheorem{example}{Example}

\newtheorem{proposition}{Proposition}

\title{The component-wise egalitarian Myerson value for Network Games}
\author{Surajit Borkotokey\footnote{Department of Mathematics, Dibrugarh University, Dibrugarh-786004, Assam, India; Email:surajitbor@gmail.com}~~ Sujata Goala\footnote{Department of Mathematics, Dibrugarh University, Dibrugarh-786004, Assam, India; Email:sujatagoala10@gmail.com}~~ Niharika Kakoty\footnote{Department of Mathematics, Dibrugarh University, Dibrugarh-786004, Assam, India; Email:nikskakoty@gmail.com}~~ Parishmita Boruah\footnote{Department of Mathematics, Dibrugarh University, Dibrugarh-786004, Assam, India; Email:pbaruahm@gmail.com}}
\date{}
\begin{document}
\maketitle
\begin{abstract}
\noindent We introduce the component-wise egalitarian Myerson value for network games. This new value being a convex combination of the Myerson value and the component-wise equal division rule is a player-based allocation rule. In network games under the cooperative framework, the Myerson value is an extreme example of marginalism, while the equal division rule signifies egalitarianism. In the proposed component-wise egalitarian Myerson value, a convexity parameter combines these two attributes and determines the degree of solidarity to the players. Here, by solidarity, we mean the mutual support or compensation among the players in a network. We provide three axiomatic characterizations of the value. Further, we propose an implementation mechanism for the component-wise egalitarian Myerson value under subgame perfect Nash equilibrium.
\end{abstract}

\section{Introduction}
We introduce the notion of a component-wise egalitarian Myerson value for network games under cooperative game-theoretic set-up due to \cite{Jackson1996}. The component-wise egalitarian Myerson value takes care of both marginalism and egalitarianism in allocating the total resource generated by a network to the players. Marginalism accounts for players' individual productivities, while egalitarianism does not distinguish between the relative efficiency of players in generating a resource. Over the past few years, network games under cooperative set-up (network games in short) have been effectively used to study various socio-economic issues related to trading and exchange, collaboration among acquaintances, security in the cyber world, sharing of natural resources among stack-holders, multi-graph problems pre-dominant in the field of information and system sciences, etc., to name a few. In a network game, the players/agents link themselves under some binding agreements and generate a worth. The problem then lies in finding a suitable allocation rule that distributes the total worth generated among the players. { The study of network games has its origin in the seminal work of \cite{Myerson} on graph restricted games in 1977. In graph-restricted games, later known as communication situations ~\cite{Jackson2005}, coalitions in which players are connected through a network are the ones who can generate non-zero worth. The structure of the network is not important in a communication situation. However, when the cost of network formation matters which may be the case, for example, of establishing physical networks among nodes, it becomes important to consider the direct and indirect links among players in the generation of the worth. Consequently, \cite{Jackson1996} proposes the network games as an extension to the communication situations where the worth of a coalition depends on the network structure through which the players in that coalition are connected. Consequently, the Myerson value first proposed as an allocation rule in \cite{Myerson} for communication situations is further extended to network games in~\cite{Jackson1996, Jackson2005}. The position value as an alternative allocation rule is proposed by \cite{Meesen} and characterized by ~\cite{Borm,Slikker05a,Slikker05b,slikker07} for communication situations and later extended to network games ~\cite{Jackson1996,Jackson2005,Nouweland}. While the position value emphasizes the links among the players more, the Myerson value concerns more on the players. Both these values are based on the productivity of the links and players, respectively; we call this: the marginalistic approach, see~\cite{manuel}.} A third allocation rule is the equal division rule which assigns equal pay-off to all the players in a network. The component-wise equal division rule assigns equal pay-offs to the players in a component. Both the equal division rule and the component-wise equal division rule are attributed to the notion of egalitarianism.  For a comprehensive study on these allocation rules, we recommend~\cite{borkotokey_rev,borkotokey_rev_1,Caulier,Herings,Jackson2005}. { The marginalistic allocation rules are insensitive to non-productive players. They do not provide incentives to players who are not currently productive but have possibilities of being productive in future collaborations. On the other hand, egalitarianism does not distinguish the players based on their productivity.  It encourages free-riding activities in a game \cite{bowels}. Therefore, in many practical situations, none of these extreme approaches can be deemed suitable for designing an allocation rule. Rules that exhibit solidarity\footnote{By solidarity we mean the mutual support or compensation among the players in a network.} to non-productive players on the one hand, and encourage the productive players to generate more worth, on the other hand, should ideally combine both these marginal and egalitarian approaches. We call them solidarity values. Take, for example, the International North-South Transport Corridor (INSTC) that connects India with Iran and other parts of Central Asia, Russia and has the scope of expansion up to the Baltic, Nordic, and Arctic regions. The trade between India and the landlocked Central Asian Region is not encouraging. The absence of an efficient transportation network with a facility for storage hubs at key nodes is identified to be the main reason for this shortfall. However, with the opening of Iran's Chabahar Port, India is expected to do well with its exports.\footnote{India's Export Opportunities Along the International North-South Transport Corridor--Naina Bhardwaj, India Briefing, June 25, 2021} Countries like Afghanistan, Turkmenistan, Bulgaria, etc., can allow their territories for establishing storage hubs and thus, contribute to increasing trades through the INSTC without being directly involved in the worth generation process. Thus, for sustaining the network among the worth generating countries, those non-productive countries should also be given a share of the total worth for their service. This we call their solidarity share. Allocation rules based on marginalism only cannot explain such solidarity-based models. Take another example involving the universal basic income proposed and piloted in many Scandinavian countries and parts of the United Kingdom.\footnote{Is Finland's basic universal income a solution to automation, fewer jobs, and lower wages?-- Sonia Sodha, The Guardian, February 19, 2017.}  The universal basic income is an unconditional income paid by the government to all the citizens, irrespective of whether they produce some worth or not. This is an example of solidarity shown to the non-productive players by the state. The supporters consider this as an answer to the fragmented welfare state that led to growing wage inequality, unemployment, an increasing number of temporary and part-time jobs. However, the critiques argue that such a guaranteed level of income already exists in most of the welfare schemes. The non-contributory nature of the welfare state, the subsidies to those in work, the children allowance and housing benefits, higher taxation slabs for the high waged strata are some of these measures already in place which exhibits solidarity to the non-productive and less productive players. They feel that a separate universal basic income as an egalitarian measure will overload the state's expenditure and burden to the productive and worth-generating sectors. On the other hand, a flat basic income may also lead to free-riding activities, which is again not desirable. Therefore, a quantified and simplistic approach having sound theoretical background is necessary to highlight how much solidarity the players want to show to the non-productive players. Moreover, it is also not desirable to exhibit solidarity to players when they are at par in terms of their per capita productivity. A suitable allocation rule should adhere to such intuitive considerations while allocating the worth among the players.  In section \ref{sec:2}, this idea is illustrated with a numerical example. More examples can be found in \cite{Dhrubajit}, where the authors introduce the Generalized egalitarian Shapley value in the cooperative framework that reflects the flexibility in choosing the level of marginality based on the coalition size.}\\
In this paper, we study the $\alpha$-component-wise egalitarian Myerson value or simply the $\alpha$-CEM value, which is a convex combination of the Myerson value and the component-wise equal division rule based on the convexity parameter $\alpha\in [0,1]$. The $\alpha$-component-wise egalitarian Myerson value provides a trade-off between marginalism and egalitarianism. The designer can adjust the solidarity to the players by a choice of $\alpha$ between $0$ and $1$. We provide three characterizations of the $\alpha$-component-wise egalitarian Myerson value. Our characterizations follow their counterparts in cooperative games due to \cite{Casajus,Casajus14a,ruiz,Brink}. { We also propose a non-cooperative bidding mechanism to show that the bidding process results in the $\alpha$-component-wise egalitarian Myerson value under subgame perfect Nash equilibrium. This approach highlights the bargaining framework of the proposed allocation rule and justifies its existence. Each stage of the bidding mechanism analyzes the players' strategic and solidarity concerns under a cooperative environment. We develop our mechanism based on similar works by \cite{castrillo, slikker07}.\\
The rest of the paper proceeds as follows. In section \ref{sec:2} we briefly discuss the preliminary concepts required to develop our model. Section \ref{sec:3} includes four characterizations of the component-wise egalitarian Myerson value. Section \ref{sec:5} describes the bidding mechanism for the component-wise egalitarian Myerson value and finally section \ref{sec:6} concludes.}

\section{Model Formulation} \label{sec:2} 
{   
In this section, we compile the preliminary notations, definitions, and results from \cite{belau2, belau3, borkotokey_rev, Caulier, Jackson2005} required for the development of our model.
\begin{itemize}
\item \textit{Players}
\newline Let $N= \{1, 2, ..., n\}$ be the set of players. A coalition $S$ is a subset of $N$. We use the corresponding lower case letter $s$ to denote the cardinality of coalition $S$, thus $\# S = s$.
\item \textit{Networks}
\newline Given the player set $N$ and distinct players $i,j\in N$, a link $ij$  is the pair $\{ i,j \}$ which represents an undirected relationship between $i$ and $j$. Clearly, $ij$ is equivalent to $ji$.
\par The set of all possible links with the player set $N$ denoted by $g_{N}=\{ij$ $|$ $i,j\in N$ and $i\neq j\}$ is called the complete network.
\par  A network $g$ is a subset of  $g_{N}$. The set of all possible networks on $N$ is $\mathbb G^N=\{ g \mid g\subseteq g_{N} \}$.
\par The network $g_0=\emptyset$ is the network without any links, which we refer as the empty network.
\par Let the number of links in a network $g$ be denoted by $l(g)$. Obviously, $l(g_{N})= \binom{n}{2} = \tfrac{1}{2} n(n-1)$ and $l(g_0)=0$. 
\par For every network $g\in\mathbb G^N$ and every player $i\in N$ we denote $i$'s neighbourhood in $g$ by $N_{i}(g)=\{j\in N\mid j\neq i$ and $ij\in g\}$ as the set of players with whom $i$ is directly linked in $g$.
\par The set of players in the network $g$ is denoted by $N(g)$. Also, denote by $n_{i}(g)=\#N_{i}(g)$, and $n(g)=\#N(g)$.
\par Denote by $L_{i}(g)=\{ij\in g\mid j\in N_{i}(g)\}\subseteq g$ the set of links of player $i$ in $g$.
\item \textit{Networks on subsets of players}
\newline For any $g\in\mathbb G^N$ and $S\subseteq N$, the restriction of $g$ on the coalition $S$ is denoted by $g|_{S}$ and is given by $g|_{S}=\{ ij \in g \mid i,j \in S \}$. 
For $g\subseteq g_{N}$ and $g'\subseteq g$, let $g-g'$ denote the network $g \setminus g'$. Similarly, for $g'\subseteq g_{N} \setminus g$, let $g+g'$ denote the network $g\cup g'$.
\item \textit{Paths}
\newline  A path in $g$ connecting $i$ and $j$ is a set of distinct players $\{i_{1},i_{2},\ldots,i_{p}\}\subseteq N(g)$ with $p\geqslant2$ such that $i_{1}=i$, $i_{p}=j$, and $\{i_{1}i_{2},i_{2}i_{3},\ldots,i_{p-1} i_{p}\}\subseteq g$.
\par We say $i$ and $j$ are connected to each other if a path exists between them and they are disconnected otherwise.
\item \textit{Components}
\newline The network {$h\subseteq g$} is a component of $g$ if for all { $i\in N(h)$} and  { $j\in N(h)$}, $i\neq j$, there exists a path in  {$h$} connecting $i$ and $j$ and for any {$i\in N(h) $} and $j\in N(g)$, $ij\in g$ implies {$ij\in h$}.
\par  In other words, a component is simply a maximally connected subnetwork of $g$. We denote the set of network components of the network $g$ by $C(g)$.
\item \textit{Isolated players} 
\newline The set of players that are not connected in the network $g$ are collected in the set of isolated players in $g$ denoted by $N_{0}(g)=N\setminus N(g)=\{i\in N\mid N_{i}(g)=\emptyset\}$. 
\par  Clearly, $N_0 (g_0) = N$.
\item \textit{Network games and value functions}
\newline  A network game is the pair $(N, v)$ where $N$ is the player set and $v \colon\mathbb G^N \to \mathbb R$ is a function called the characteristic function that satisfies $v (g_0)=0$.
\par  The space of all network games with player set $N$ is denoted by $\mathbb V^N$, which is a $2^{l(g)}-1$ dimensional vector space. If no ambiguity arises on the player set $N$, we denote a network game simply by its characteristic function $v$.
\item \textit{component additive network games}
\newline  A network game $v$ is called component additive if for every network $g \in \mathbb G^N$, $v(g) = \sum_{h \in C(g)} v(h)$. Component additivity ensures that there is no externality across different components in a network.
\par The class of component additive games is denoted by $\mathbb H^N$, which is a $\sum_{h \in C(g)}2^{ l(h)}-1$ dimensional subspace of  $\mathbb V^N$.
\par {Unless stated explicitly, throughout this paper, we only consider the component additive network games}. 

\item \textit{Basis for value functions}
\newline Two important subclasses of games in $\mathbb V^N$ are the class of unanimity games and the class of identity games which are defined as follows.
\begin{itemize}
\item For any network $g \in \mathbb G^N$, the unanimity game $(N, u_{g})$ is defined as: 
\[ u_{g}(g')=\begin{cases} 
        1 &  \text{if} ~ g \subseteq g' \\
        0& ~\text{otherwise}.  
    \end{cases}
 \]
 \item For any network $g \in \mathbb G^N$, the identity game $(N,e_{g})$ is defined as: 
 \[ e_{g}(g')=\begin{cases} 
         1 &  \text{if} ~ g' = g\\
          0& ~ \text{otherwise}.  
     \end{cases}
  \]
\end{itemize}
These two classes of network games are standard bases for $\mathbb V^N$.
\item \textit{Allocation rules}
\newline A network allocation rule is a function $Y\colon\mathbb G^N \times \mathbb V^N \to \mathbb R^N$ such that for every $g \in \mathbb G^N$ and every $v \in \mathbb V^N$, $Y_i(g,v)= 0$ whenever $i \in N_0(g)$.
\item \textit{Efficiency}
\newline  An allocation rule $Y$ is efficient if for each $g \in \mathbb G^N$ and $v \in \mathbb V^N$ it holds that $$\sum_{i\in N(g)} Y_i(g, v) = v(g).$$
\par  Thus, an efficient network allocation rule determines how the collective worth generated by a given network $g \in \mathbb G^N$ with respect to the network game $v \in \mathbb V^N$ is allocated to the players.
\item \textit{Component Balanced}
\newline An allocation rule $Y$ is Component Balanced if for every $v \in \mathbb H^N$ and $g \in \mathbb G^N$ it holds that $$\sum_{i\in N(h)}Y_{i}(g,v)=v(h), \;\textrm{for every component $h \in C(g)$.}$$
\item \textit{Equal Bargaining Power}
\newline The allocation rule $Y$ satisfies Equal Bargaining Power if for every network $g \in \mathbb G^N$, {component additive network game}  $v \in \mathbb H^N$, and for all  $i,j \in N(g)$, we have
\begin{equation}
	Y_{i}(g,v)-Y_{i}(g-ij,v)=Y_{j}(g,v)-Y_{j}(g-ij,v).
\end{equation}
\item \textit{Balanced Contributions Property}
\newline The allocation rule $Y$ satisfies the Balanced Contributions Property if for every network $g \in\mathbb G^N$, every {$v \in \mathbb H^N$}, and for all players $i,j \in N(g)$, we have
\begin{equation}
	Y_{i}(g,v)-Y_{i} \left( g-L_{j}(g),v \right)=Y_{j}(g,v)-Y_{j} \left( g-L_{i}(g), v \right).
\end{equation}
\item \textit{Balanced Component Contributions Property}
\newline Denote by $h_i^g$ the component of $g\in \mathbb G^N$ containing player $i\in N$ \footnote{ $h_i^g$ is empty when $i$ is an isolated player.}. The allocation rule $Y$ satisfies the Balanced Component Contributions Property if for every network $g \in \mathbb G^N$, {every  $v \in \mathbb H^N$}, and for all players $i,j \in N(g)$, we have
\begin{equation}
	Y_{i}(g,v)-Y_{i} \left( g- h_j^g,v \right)=Y_{j}(g,v)-Y_{j} \left( g-h_{i}^g,v \right),
\end{equation} 
\item \textit{The Myerson value}
\newline In the following we define the Myerson value due to ~\cite{Myerson} which is extended to network games by \cite{Jackson1996}. 
\par The Myerson value is the network allocation rule $Y^{MV} \colon \mathbb G^N \times \mathbb V^N\to \mathbb R^N$ given by
\begin{equation}\label{eq:myerson}
 Y_{i}^{MV}(g,v)={\sum_{S\subseteq N\setminus i}} \left(v(g|_{S\cup i})-v(g|_{S})\right)  \left( \frac{s!(n(g)-s -1)!}{n(g)!}\right).
\end{equation}
for every $g \in \mathbb G^N$ and $v \in \mathbb V^N$.

\item \textit{The component-wise equal division rule}
\newline The component-wise equal division rule is studied in \cite{Jackson1996,slikker07} and is the network allocation rule $Y^{CE} \colon \mathbb G^ N\times \mathbb V^N\to\mathbb R^N$ defined by
 \begin{equation}\label{eq:ed}
 Y_{i}^{CE}(g,v)=\left\{\begin{array}{ll}
      & \dfrac{v(h_i^g)}{n(h_i^g)} \;\;\;\textrm{if there exists $h_i^g \in C(g)$ such that $i\in h_i^g$,} \\
      & 0 \;\;\;\textrm{otherwise.}
     \end{array}\right.
 \end{equation}
 for every $g \in \mathbb G^N$ and $v \in \mathbb V^N$.
\end{itemize}
\begin{remark}
\par The Myerson value aggregates the marginal contributions of a player in all her networks. On the other hand, the component-wise equal division rule equally splits the total worth of a component in a network among all the members of the component. On $\mathbb H^N$, both Myerson value and the component-wise equal division rule are Component Balanced and Efficient. \cite{Jackson1996} shows that the Myerson value is the unique network allocation rule satisfying both Component Balance and Equal Bargaining Power on the class of component additive games.  \cite{slikker07} gives an alternative characterization of the Myerson value where the Equal Bargaining Power is replaced by Balanced Contribution Property.  \cite{slikker07} also characterizes the component-wise equal division rule by Component Balance and Balanced Component Contributions.
\end{remark}
}
\begin{itemize}
 \item \textit{The component-wise egalitarian Myerson value}
 \newline For $\alpha \in [0,1]$, let the $\alpha$-component-wise egalitarian Myerson value be denoted by $Y^{\alpha-CEM}\colon \mathbb G^N \times \mathbb V^N\to\mathbb R^N$ and defined for every $g \in \mathbb G^N$ and $v \in \mathbb V^N$ by
\begin{equation}\label{emvalue}
Y^{\alpha-CEM}(g,v) = \alpha Y^{MV}(g, v) + (1-\alpha)Y^{CE}(g,v).
\end{equation}
\end{itemize}     
{
\begin{remark}
The $\alpha$-component-wise egalitarian Myerson value is an intuitive solidarity value in the sense that while it exhibits solidarity to the weaker or non-productive players in the game, it also distances away from being un-necessarily altruistic. To see this, consider the following two examples.
\end{remark} 
\begin{example}
Take $N = \{1,2,3,4,5\}$. Consider the network $g=\{12,34,45\}$ and a value function $v$ given by $v(\{12\})=v(\{45\})=v(\{34,45\})=v(\{12,34\})=2$, $v(\{12,45\})=v(\{12,34,45\})=4$, and $v(g')=0$ for all other  $g' \in \mathbb{G}^{N}$. \\
The network $g$ looks like as in Figure~{\color{red} 1}. Here, player $3$ is a non-productive player\footnote{We will call this as the superfluous player later.}. The Myerson value $Y^{MV}_{i}(g,v)$, component-wise equal division rule $Y^{CE}_{i}(g,v)$, and the component-wise egalitarian Myerson value $Y^{\alpha -CEM}_{i}(g,v)$ for $i \in \{1,2,3,4,5\}$ and $\alpha\in [0,1]$ are given in Table~{\color{red} 1}. 
\end{example}
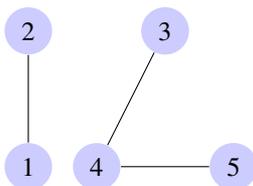
\begin{figure}[h!]\label{fig-1}
\begin{center}
\begin{tikzpicture}  
  [scale=.9,auto=center,every node/.style={circle,fill=blue!20}]   \node (a1) at (2,2) {1};  
  \node (a2) at (2,4)  {2};   
  \node (a3) at (4,4)  {3};  
  \node (a4) at (3,2) {4};  
  \node (a5) at (5,2)  {5}; 
  \draw (a1) -- (a2); 
  \draw (a4) -- (a5);  
  \draw (a3) -- (a4);  
\end{tikzpicture} 
\end{center}
\caption{Network Game 1}
\end{figure}
\begin{table}[h!]\label{table-2}
\begin{center}
\begin{tabular}{|c|c|c|c|}
\hline
   $i$  &  $Y^{MV}_{i}(g,v)$ &  $Y^{CE}_{i}(g,v)$ & $Y^{\alpha -CEM}_{i}(g,v)$\\
   \hline
    1 &  1 & 1 & 1 \\
    \hline
    2 &  1 & 1 & 1 \\
    \hline
    3 &  0 & $\frac{2}{3}$ & $\alpha .0+(1-\alpha).\frac{2}{3}$ \\
    \hline
    4 &  1 & $\frac{2}{3}$ & $\alpha .1+(1-\alpha).\frac{2}{3}$ \\
    \hline
     5 &  1 & $\frac{2}{3}$ & $\alpha .1+(1-\alpha).\frac{2}{3}$ \\
     \hline
\end{tabular}
\end{center}
\caption{Network Game 1- Players with different level of  productivity. }
\end{table}
\noindent Observe that here, player $1$ and $2$'s per capita productivities are equal and independent to the other component i.e., $\{34,45\}$. The component-wise egalitarian Myerson values $Y^{\alpha -CEM}(g,v)$ assign identical payoff to both $1$ and $2$ across all values of $\alpha$. Since player $3$ is non-productive, hence, $Y^{MV}_{3}(g,v)=0$. But the component-wise egalitarian Myerson values $Y^{\alpha -CEM}(g,v)$ exhibits solidarity to player $3$ and, therefore, she gets some positive payoffs determined by the parameter $\alpha$. 

\noindent Figure~{\color{red}2} illustrates the distribution of the payoffs to the players given by the component-wise egalitarian Myerson values $Y^{\alpha -CEM}(g,v)$ for $\alpha \in [0,1]$.

\begin{figure}[h!]\label{FIG-2}
\begin{center}
\includegraphics[width=10cm, height=6cm]{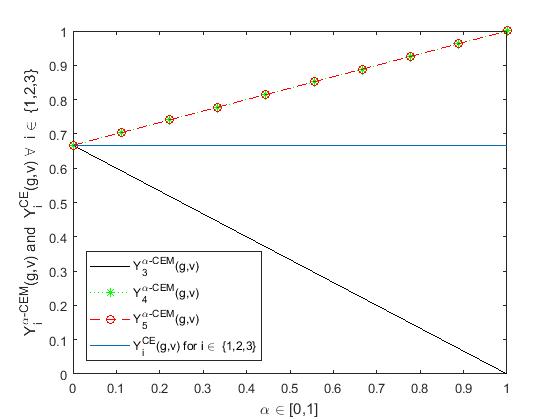}\\
\caption{Network Game 1 - Payoff distribution when players are not equally productive}
\end{center}
\end{figure}

Next, consider the following example. 

\begin{example}
Let $N =\{1,2,3,4,5\}$. Consider the network $g=\{12,34,45,35\}$ on $N$ and an additive value function 
\begin{equation*}
v=  2 u_{\{12\}}+ 3 u_{\{34,45,35\}}.
\end{equation*}
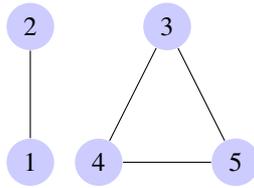
\begin{figure}[h!]\label{fig-11}
\begin{center}
\begin{tikzpicture}
 [scale=.9,auto=center,every node/.style={circle,fill=blue!20}]   \node (a1) at (2,2) {1};  
  \node (a2) at (2,4)  {2};   
  \node (a3) at (4,4)  {3};  
  \node (a4) at (3,2) {4};  
  \node (a5) at (5,2)  {5}; 
  \draw (a1) -- (a2); 
  \draw (a4) -- (a5);  
  \draw (a3) -- (a4);  
  \draw (a3) -- (a5);  
\end{tikzpicture} 
\caption{Network Game 2}
\end{center}
\end{figure}

Figure {\color{red} 3} shows that $g$ has two sub-networks through which the players interact. The productivities of each sub-network is independent. Each player's per capita productivity from her subnetwork is the same, namely $1$. Therefore, there is neither a need to exhibit solidarity within the sub-networks nor outside the sub-networks, i.e., players should be rewarded according to their common individual productivity, which is $1$. This is a reasonable property of an allocation rule intended to reflect solidarity. The component-wise egalitarian Myerson value gives all the players $1$ following this property. This makes it a reasonable solidarity value for network games. This, we show in Figure~{\color{red} 4}. 

\begin{figure}[h!]\label{fig_4}       
\begin{center}
\includegraphics[width=10cm, height=6.6cm]{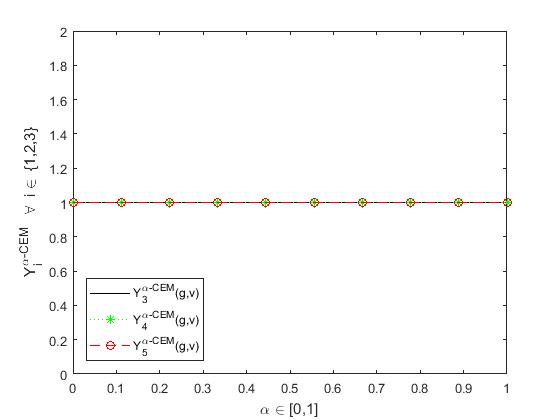}\\
\caption{ Network Game 2 - Payoff distribution when players are equally productive}
\end{center}      
\end{figure} 
\end{example}

In the next section, we characterize the component-wise egalitarian Myerson value. }
\section{Characterization}\label{sec:3} 
Before the main results, we list the following axioms, some of which are standard in cooperative game theory and are trivial extensions to their network counterparts, and some are specific to network games. We assume that there are no externalities in worth generation across components in a network and, therefore, from now onwards, unless otherwise specified, we consider only the class $\mathbb H^N$ of component additive games.
\begin{axiom}\rm \textbf{Linearity}: \label{axiom3}
For each pair of {network games} $v, w \in \mathbb H^N$ and $g\in \mathbb G^N$, 
\begin{equation}\label{eq:lin}
Y_{i}(g, av + bw)=aY_{i}(g, v)+ b Y_i(g, w)~~\forall i\in N(g)\;\textrm{and every pair $a, b \in \mathbb R$.}
\end{equation}  
If Eq.(\ref{eq:lin}) holds only for $a = b= 1$, that is, if we have 
$Y_{i}(g, v+w)=Y_{i}(g, v)+Y_{i}(g, w)~~\forall i\in N(g),$ then $Y$ is said to satisfy additivity.
\end{axiom}
\noindent Let $\pi : N\rightarrow N$ be a permutation. For $v \in \mathbb V^N$, define the network game $\pi v : \mathbb G^N \rightarrow \mathbb R$ such that $\pi v(\pi g) = v(g)$ where the network $\pi g$ is defined as $\pi g = \{ij|\; i,j \in N \;\textrm{such that}\; \pi^{-1} (i) \pi^{-1} (j) \in g\}$. It follows that if $v$ is component additive then so is $\pi v$. Therefore, there is no loss of generality in our earlier assumption of taking only component additive network games in our model. Moreover, we have for each $S \subseteq N$ and permutation $\pi:N \rightarrow N$, $\pi v(\pi g|_{\pi S})= v(g|_S)$. The next axiom follows.  
\begin{axiom}\rm \textbf{Anonymity}:\label{axiom6}
Allocation rule $Y$ satisfies Anonymity namely, $Y_{i}(g, v)=Y_{\pi(i)}(\pi g,\pi v)$, for all permutations $\pi$, $v \in \mathbb H^N$, and $g \in \mathbb G^N$ where $\pi v$ is defined as above. 
\end{axiom}
\begin{definition}
Player $i \in N(g)$ is called a superfluous player in $v \in \mathbb H^N$ if $v(g)= v(g-L_{i}(g))$ for all $g\in \mathbb G^N$. 
\end{definition}
\noindent {The superfluous player in a network game is analogous to the null player in cooperative games as removal of the superfluous player does not affect the generation of the worth by the network. However, they differ in the sense that unlike the null player in cooperative games, removal of any player from a network makes all her links in that network redundant, and the communication through all these links is lost. Therefore, the requirement of a player to be superfluous is much stronger than that of a null player. The following axiom has its counterpart in cooperative games under the name of ``null player in a productive environment" due to \cite{Casajus}.    }
\begin{axiom}\rm \textbf{Superfluous Player in a Productive Network:}\label{axiom9} 
For each network game $v\in \mathbb H^N$, and the superfluous player $i$ in $v$ and for every $g\in \mathbb G^N$ we have, $v(g)= v(g-L_{i}(g))\geq 0 $ implies $Y_{i}(g,v)\geq0.$
\end{axiom}
\noindent The Superfluous Player in a Productive Network Property ensures that the superfluous players should not be penalized for being non-productive in a game. Thus, this axiom shows solidarity to the superfluous players.
{
\begin{remark}
Note that the Anonymity and Superfluous Player in a Productive Network are specific to players in the network and, therefore, they can be considered player-based axioms. However, they differ from their counterparts in cooperative games as in their definition; the inherent network structure plays a crucial role. Removing a player from a coalition is different from removing a player from the network since the removal of the player from a network implies the removal of all the links this player makes with the other players in the network. 
\end{remark}
}
{\begin{axiom} \textbf{Component-wise Weak Monotonicity}: \label{axiom13}
Let   $g \in \mathbb G^N$ be fixed, $i \in N(g)$. Let $h \in C(g)$ be such that  $i \in N(h)$. Suppose $\pi: N \mapsto N$ be a permutation such that $\pi(k)=k$ for all $k \in N \setminus N(h)$. Let $v,w \in \mathbb H^N$ which satisfy the following two conditions:
\begin{enumerate}[~~~(i)]
\item $v(h) \geq w(h)$
\item $v(g|_{S})-v(g|_{S \setminus i})\geq w(\pi g|_{S})-w(\pi g|_{S \setminus i}),~~ \forall S \subseteq N(h)~ \text{with} ~ i \in S.$
\end{enumerate}
Then we have $Y_{i}(g, v)\geq Y_{i}(\pi g , w)$.
\end{axiom}}
{\begin{axiom} \textbf{Component-wise Local Monotonicity}:\label{axiom14}
For all $g \in \mathbb G^N$, $v \in \mathbb H^N$, and $i,j \in N(h)$ where $h \in C(g)$, if $v(g|_{S \cup i})\geq v(g|_{S \cup j}),~~ \forall S \subseteq N(h) \setminus\{i,j\}$, then we have $Y_{i}(g,v) \geq Y_{j}(g, v)$. 
\end{axiom}}
\begin{remark}\label{rem:3}
For all $g \in \mathbb G^N$, $v \in \mathbb H^N$, and $i,j \in N(h)$ where $h \in C(g)$, if $v(g|_{S \cup i})= v(g|_{S \cup j}),~~ \forall S \subseteq N(h) \setminus\{i,j\}$ and $Y$ satisfies Component-wise Local Monotonicity, then we have $Y_{i}(g,v) = Y_{j}(g, v)$ by repeated application of the inequality from both sides.
\end{remark}
{\begin{axiom} \rm \textbf{Component-wise Strong Differential Monotonicity}: \label{axiom15}
For each $v,w \in \mathbb H^N$, $g \in \mathbb G^N$, $i,j \in N(h)$ for component $h \in C(g)$, and $S\subseteq N(h) \setminus \{i,j\}$, if $v(g|_{S\cup i})- v(g|_{S\cup j})\geq w(g|_{S\cup i})- w(g|_{S\cup j})$, then we have $Y_{i}(g,v)-Y_{j}(g,v)\geq Y_{i}(g,w)-Y_{j}(g,w)$.
\end{axiom}}
\noindent The last three axioms have their origin in cooperative games due to \cite{Casajus,Casajus14b,Brink}. They reconcile monotonicity with egalitarianism and are useful for providing alternative characterizations of the component-wise egalitarian Myerson value. \\  
The following proposition is required for our first characterization.
\begin{proposition}\rm \label{prp3}
If the allocation rule $Y: \mathbb G^N \times \mathbb H^N \rightarrow \mathbb R^N$ satisfies Component-wise Weak Monotonicity and Anonymity, then it also satisfies Component-wise Local Monotonicity. 
\end{proposition}
\begin{proof}
Let $Y$ satisfy Component-wise Weak Monotonicity and Anonymity. Let $v \in \mathbb H^N$ and $g \in \mathbb G^N$ be such that, for {$i,j \in N(h)$ and an arbitrary  component $h \in C(g)$}, we have
\begin{align}\label{eq21}
v(g|_{S \cup i})\geq v(g|_{S \cup j})~~\forall S \subseteq {N(h) \setminus\{i,j\}}.\tag{21}
\end{align}
Consider a permutation $\pi : N \rightarrow  N $ such that $\pi(i)=j,\pi(j)=i,\pi(k)=k $ for $k \in N\setminus\{i,j\}$. Also take the two games $v, \pi v \in \mathbb H^N$. The following two cases arise {for each $S \subseteq N(h) \setminus i$}.\\ 
\textbf{Case~(a)}:
{$S \subseteq N(h) \setminus \{i,j\}$}. From Anonymity and Eq.(\ref{eq21}), we obtain 
\begin{eqnarray*}
[v(g|_{S \cup i}) \;-\; v(g|_{S})] -[\pi v(\pi g|_{S \cup i}) - \pi v(\pi g|_{S})]~~~~~~~~~~~~~~~~~~~~~~~~~~~~~~~~~~\\
=[v(g|_{S \cup i}) - v(g|_{S})] -[\pi v(\pi g|_{{\pi(S) \cup \pi (j)}}) - \pi v(\pi g|_{\pi (S)})]\\
= v(g|_{S \cup i}) - v(g|_{S})- v(g|_{S \cup j})+v(g|_{S})~~~~~~~~~~~~~~~~~~~~~~~~~~~~\\
= v(g|_{S \cup i})- v(g|_{S \cup j}) \geq 0.~~~~~~~~~~~~~~~~~~~~~~~~~~~~~~~~~~~~~~~~~~~~~~~~~~~ 
\end{eqnarray*}
\textbf{Case (b)}:  Let {$S \subseteq N(h) \setminus i$} be such that $j \in S$. Thus, we can write  $S= T \cup j$ where {$T \subseteq N(h) \setminus \{i,j\}$}. From Anonymity and Eq.(\ref{eq21}) again, we get
\begin{eqnarray*}
{[v(g|_{S \cup i})-v(g|_{S})]}-{[\pi v(\pi g|_{S \cup i})- \pi v(\pi g|_{ S})]}~~~~~~~~~~~~~~~~~~~~~~~~~~~~~~~~~~~~~~~~~~~~~~~~~~~~~~~~~~~~~~~~~~~~~~~\\
=[v(g|_{T \cup i\cup j})-v(g|_{T \cup j})]-{[ \pi v(\pi g|_{\pi (T) \cup \pi(i)\cup \pi(j))})-\pi v(\pi g|_{\pi (T) \cup \pi(i)})]}~~~~\\
= {v(g|_{T \cup i\cup j}) - v(g|_{T \cup j})-  v( g|_{T \cup i \cup j})+  v(g|_{T \cup i})}~~~~~~~~~~~~~~~~~~~~~~~~~~~~~~~~~~~~~~~~~~\\
= {v(g|_{T \cup i})- v(g|_{T \cup j}) \geq 0.}~~~~~~~~~~~~~~~~~~~~~~~~~~~~~~~~~~~~~~~~~~~~~~~~~~~~~~~~~~~~~~~~~~~~~~~~~~~~~~~
\end{eqnarray*}
Moreover, $v(g)=\pi v(\pi g)$, hence the pair $v$ and $\pi v$ satisfy the conditions of Component-wise Weak Monotonicity. Thus, combining with Anonymity, Component-wise Weak Monotonicity gives,
\begin{eqnarray*}
Y_{i}(g,v) &\geq& Y_{i}(\pi g, \pi v)\\
&=& Y_{\pi (j)}(\pi g, \pi v)\\
&=&Y_{j}( g, v)\\
\Rightarrow Y_{i}(g,v) &\geq&  Y_{j}( g, v).
\end{eqnarray*}
This completes the proof. 
\end{proof}
\begin{theorem}\rm \label{thm3}
A network allocation rule $Y:\mathbb G^N \times \mathbb H^N \mapsto \mathbb R^N$ is Component Balanced and satisfies Anonymity, Linearity, and Component-wise Weak Monotonicity if and only if it is the $\alpha$-component-wise egalitarian Myerson value $Y^{\alpha-CEM}$ for some $\alpha \in [0,1]$.
\end{theorem}
\begin{proof}
It is easy to show that the $Y^{\alpha-CEM}$ is Component Balanced and satisfies Anonymity, Linearity, and Component-wise Weak Monotonicity. \\
For the uniqueness, assume  that an allocation rule $Y$ satisfy these axioms.\\ 
Let $v_{0}$ be the null game given by $v_{0}(g)=0 \;\forall g \in \mathbb G^N.$ \\
Then, by Component Balance and Anonymity, 
\begin{equation}\label{eq:null_game}
Y_{i}(g, v_{0})=0= Y_{i}^{\alpha-CEM}(g, v_{0}), ~~\forall i \in N(g). 
\end{equation}
This we call the null game property.\\
Note that, each $v\in \mathbb H^N$ such that $v \ne v_0$, can be represented as
\begin{equation}\label{eq:1}
v=\sum_{g' \in \mathbb G^N}\Delta_{g'}(v)u_{g'}
\end{equation} 
where $\Delta_{g'}(v)$ is the unique unanimity coefficient corresponding to the unanimity game $u_{g'}$ called the Harsanyi dividend\footnote{for more details, see~\cite{Nouweland}}. Thus, by Linearity, it suffices to show the uniqueness of the $\alpha$-component-wise egalitarian Myerson value on the unanimity games. Moreover, for all $v \in  \mathbb H^N$, $\Delta_{g'}(v)=0$, if $g' \in \mathbb G^N $ contains links from two different components\footnote{This result is proved in \cite{Nouweland} (Lemma 1, page 267).}. Hence, we require to show that $Y_{i}(g, u_{g'})=Y^{\alpha-CEM}_{i}(g, u_{g'})$, only for connected $g' \in \mathbb G^N$. Thus, for $g', g \in  \mathbb G^N$, we have the following three cases: \\
\textbf{Case~(a)}: $g' \subseteq g$. First, we assume that $g$ is connected. We use the method of induction on $n(g')\geq 2$. 
\\
Let $n(g')=2$, i.e., $g'=\{ij\}$ for $i,j \in N$. Consider the permutation $\pi : N \rightarrow  N$ such that $\pi(i)=j$, $\pi(j)=i$  and $\pi(k)=k$ for all {$k \in N(g) \setminus\{i,j\}$}.\\
{For $S \subseteq N(g) \setminus\{i,j\}$, we have $g' \not\subseteq g|_{S \cup i}$ and $g'\not\subseteq g|_{S \cup j}$.} So, by Anonymity, 
\begin{eqnarray*}
[u_{g'}(g|_{S \cup i})-u_{g'}(g|_{S})]&-&[\pi u_{g'}(\pi g|_{S \cup i})-\pi u_{g'}(\pi g|_{S})]\\
&=& [u_{g'}(g|_{S \cup i})-u_{g'}(g|_{S})]-{[\pi u_{g'}(\pi g|_{\pi (S) \cup \pi (j)})-\pi u_{g'}(\pi g|_{\pi (S)})]}\\
&=& u_g'(g|_{S \cup i}) - u_g'(g|_{S \cup j})\\
&=&0 \;\;
\end{eqnarray*}
Again, { for $S \subseteq N(g) \setminus i$,} let us take $S=T \cup j$ where $T \subseteq N(g) \setminus\{i,j\}$. Then, using Anonymity again, we have
\[[u_{g'}(g|_{S \cup i})-u_{g'}(g|_{S})]-[\pi u_{g'}(\pi g|_{S \cup i})-\pi u_{g'}(\pi g|_{ S })]~~~~~~~~~~~~~~~~~~~~~~~~~~~~~~~~~~~~~~~~~~~~~~~~~~~~~~~~~~~~~~~~\]
\begin{eqnarray*}
&=&[u_{g'}(g|_{T \cup i \cup j})-u_{g'}(g|_{T \cup j})]-[\pi u_{g'}(\pi g|_{ T \cup i \cup j})-\pi u_{g'}(\pi g|_{T \cup j })]\\
&= &[u_{g'}(g|_{T \cup j \cup i})-u_{g'}(g|_{T \cup j})]-{[\pi u_{g'}(\pi g|_{\pi { (T) \cup \pi(j) \cup \pi(i)}})-\pi u_{g'}(\pi g|_{\pi (T) \cup \pi(i)})]}\\
&= &[u_{g'}(g|_{T \cup j \cup i})-u_{g'}(g|_{T \cup j})]-[ u_{g'}( g|_{T \cup i \cup j})- u_{g'}(g|_{T \cup i})]\\
&= &{u_{g'}(g|_{T \cup i})- u_{g'}(g|_{T \cup j})}=0\;\; 
\end{eqnarray*}
By definition, we have $u_{g'}(g)=\pi u_{g'}(\pi g)$. Hence, by Component-wise Weak Monotonicity and Anonymity, we get, 
\begin{equation}\label{eq:10}
Y_{i}(g,u_{g'})\geq Y_{i}(\pi g,\pi u_{g'})={Y_{\pi (j)}}(\pi g,\pi u_{g'}) =Y_{j}(g,u_{g'}).
\end{equation}
However, $i$ and $j$ are independent of one another and, therefore, we also have 
\begin{equation}\label{eq:11}
Y_{j}(g,u_{g'}) \geq Y_{i}(g,u_{g'}).
\end{equation}
Thus, combining Eq.(\ref{eq:10}) and Eq.(\ref{eq:11}), we obtain $Y_{i}(g,u_{g'})= Y_{j}(g,u_{g'})$. 
Now, take any arbitrary pair $k,l \in N(g) \setminus N(g')$, and consider the permutation $\pi : N \rightarrow  N$ such that $\pi(k)=l$ and $\pi(l)=k$ and $\pi(i)=i$ for all $i \in N \setminus\{k,l\}$. Then, there exists a $\beta^{*} \in \mathbb{R}$ such that 
\begin{equation}
Y_{k}(g,u_{g'})=Y_{l}(g,u_{g'})=\beta^{*}.
\end{equation}
Now, $k$ and $l \in N(g)\setminus \{k,i,j\}$ being arbitrary, we finally obtain 
\begin{equation}
Y_{k}(g,u_{g'})=\beta^{*}\;\;\textrm{for all $k \in N(g)\setminus \{i,j\} $}.
\end{equation}
$Y$ being Component Balanced, we have 
\begin{equation}\label{Eq:0}
\sum_{i \in N(g)}Y_{i}(g,u_{g'})=u_{g'}(g)
\end{equation}
This would imply 
\begin{equation} 
2Y_{i}(g,u_{g'})+(n(g)-2)\beta^{*}=1,
\end{equation} that is, 
\begin{equation}
Y_{i}(g,u_{g'})=\dfrac{1-(n(g)-2)\beta^{*}}{2}.
\end{equation}
It follows that, 
\begin{equation}
Y_{i}(g,u_{g'})=\beta^{*}+\dfrac{1-n(g) \beta^{*}}{2}.
\end{equation}
Again, recall from proposition (\ref{prp3}) that $Y$ satisfies Component-wise Local Monotonicity. \\
Further, for $i \in N(g), k \in N(g)\setminus N(g')$ we must have,  
\begin{equation}
u_{g'}(g|_{S \cup i}) \geq u_{g'}(g|_{S  \cup k})\;\;\textrm{whenever $S \subseteq N \setminus \{i,k\}$}.
\end{equation}
Hence,  by Component-wise Local Monotonicity, it holds that 
 \begin{equation}\label{Eq:1} Y_{i}(g,u_{g'}) \geq Y_{k}(g,u_{g'})=\beta^{*}.
 \end{equation}
Now, consider the two games $u_g'$ and $v_0$ and use Eq.(\ref{eq:null_game}) along with Component-wise Weak Monotonicity and obtain 
\begin{equation}\label{Eq:2} 
Y_{k}(g,u_{g'})\geq Y_{k}(g,v_{0})=0 \Rightarrow  \beta^{*} \geq 0.
\end{equation}
From Eq.(\ref{Eq:0}), Eq.(\ref{Eq:1}), and Eq.(\ref{Eq:2}), we obtain 
\begin{equation}\label{eq:21}
1-n(g) \beta^{*} \geq 0\;\;\textrm{and $\beta^{*} \geq 0$}.
\end{equation}
Therefore, there exists an $\alpha\in [0,1]$ such that $ \beta^{*}=\dfrac{1-\alpha}{n(g)}$ and consequently, 
\begin{equation}\label{eq:22}
Y_{i}(g,u_{g'})=Y_{j}(g,u_{g'})=\dfrac{\alpha}{2}+\dfrac{1-\alpha}{n(g)}.
\end{equation}
Therefore, we finally obtain,
\begin{equation}\label{SPPN1}
 Y_{i}(g,u_{g'}) =\begin{cases} 
  \dfrac{1-\alpha}{n(g)}+\dfrac{\alpha}{2} & \text{if} ~i\in N(g') \\
   \dfrac{1-\alpha}{n(g)} & \text{if} ~i \in  N(g) \setminus  N(g')\\
   0 & \text{otherwise}  
    \end{cases}
\end{equation}
\noindent Since $g$ is connected, we have for $k \in N(g')$, 
\begin{equation}     
Y^{MV}_{i}(g,u_{g'})=\dfrac{1}{2}=Y^{MV}_{j}(g,u_{g'}).
\end{equation} 
and whenever $k \in N(g) \setminus N(g')$ we have, 
\begin{equation}
Y^{MV}_{k}(g,u_{g'})=0.
\end{equation}
Also, 
\begin{equation}
Y^{CE}_{i}(g,u_{g'})=\dfrac{1}{n(g)}\;\; \textrm{for all $i \in N(g)$.}
\end{equation} 
It follows that, 
\begin{align}
Y_{i}(g,u_{g'})&=(1-\alpha)Y^{CE}_{i}(g,u_{g'})+\alpha Y^{MV}_{i}(g,u_{g'})=Y^{\alpha-CEM}_{i}(g,u_{g'}), ~\forall ~ i \in N(g).
 \end{align}
\noindent Now, take $u_{g'}$ such that $2 < n(g') < n(g) $ and let $Y$ satisfy the given axioms. \\
Let $j \in N(g)\setminus N(g')$. By Component-wise Weak Monotonicity and the induction hypothesis we get,
\begin{align}\label{eq:13}
Y_{j}(g,u_{g'})&=Y_{j}(g,u_{(g'- L_i(g))})=Y^{\alpha-CEM}_{j}(g,u_{(g'- L_i(g))})=\dfrac{1-\alpha}{n(g)}={Y_{j}^{\alpha-CEM}(g,u_{g'})}.
\end{align}
Since $Y$ is Component Balanced, we have,
\begin{align}\label{eq:14}
\sum_{i \in N(g')}Y_{i}(g,u_{g'})&=u_{g'}(g)-\sum_{i \in N(g)\setminus  N(g')}Y_{i}(g,u_{g'})=1-[n(g)-n(g')]\times\dfrac{1-\alpha}{n(g)}
\end{align}
Again by Anonymity,
\begin{align}
Y_{i}(g,u_{g'})&=\dfrac{1-[n(g)-n(g')]\times\dfrac{1-\alpha}{n(g)}}{n(g')}\nonumber \\
&=\dfrac{1-\alpha}{n(g)}+\dfrac{\alpha}{n(g')},~\text{for all}~ i\in N(g')\nonumber \\
&=(1-\alpha)Y_{i}^{CE}(g,u_{g'})+\alpha Y_{i}^{{MV}}(g,u_{g'})\nonumber \\
&={Y_{i}^{\alpha-CEM}(g,u_{g'})}.\label{eq:15}
\end{align}
Hence, using Eq.(\ref{eq:13}), Eq.(\ref{eq:14}), and Eq.(\ref{eq:15}) we get,
\begin{equation*}
Y_{i}(g,u_{g'})=\left\{\begin{array}{ll}
               & Y_{i}^{\alpha-CEM}(g,u_{g'}) \;\;\textrm{for all $i \in N(g)$ such that $g' \subseteq g $}\\                
               & 0\;\;\textrm{otherwise.}
                \end{array}\right.
\end{equation*}
Now, take $g' \subseteq g$ where $g$ contains more than one component. Without loss of generality we assume that $g=g_{1} + g_{2}$ and $g_1 \cap g_2 =\emptyset$. Moreover, since $g'$ is connected, therefore, either we have $g' \subseteq g_{1}$ or $g' \subseteq g_{2}$. {Let without loss of generality, $g' \subseteq g_{1}$}. Since $u_{g'}$ is also component additive\footnote{This result is proved in \cite{Nouweland} (Corollary 1 of Lemma 1, page 268).}, therefore $u_{g'}(g_{1} + g_{2})=u_{g'}(g_{1} )+ u_{g'}( g_{2})=1$. Then, using Component Balance, Anonymity, and Component-wise Weak Monotonicity of $Y$ we get
\begin{eqnarray*}
\sum_{i \in N(g_{_1}) \cup N(g_{_2} )} Y_{i}(g,u_{g'})=1
&\Rightarrow & \sum_{i \in N(g_{_1} )} Y_{i}(g,u_{g'})+\sum_{i \in  N(g_{_2} )} Y_{i}(g,u_{g'})=1.\\
&\Rightarrow & 1 + n(g_{2}) Y_{i}(g,u_{g'}) = 1 \;\;\forall i \in N(g_2).\\
&\Rightarrow &  n(g_{2}) Y_{i}(g,u_{g'})=0\;\;\forall i \in N(g_2).
\end{eqnarray*} 
\noindent Hence, 
\begin{equation}
Y_{i}(g,u_{g'})=0= {Y_{i}^{\alpha -CEM}(g,u_{g'})} \;\;\textrm{for all $i \in N(g_{2})$ with $N(g') \cap N(g_{2})=\emptyset$.}
\end{equation}
Moreover, under the same set of axioms i.e., Component Balance, Anonymity, and Component-wise Weak Monotonicity of $Y$, we obtain
\begin{eqnarray}
\sum_{i \in N(g_{_1} )} Y_{i}(g,u_{g'})=1.~~~~~~~~~~~~~~~~~~~~~~~~~~~~~~~~~~~~~~~~~&&\nonumber\\
\Rightarrow \sum_{k \in N(g_{_1}) \setminus N(g')}Y_{i}(g,u_{g'})+ \sum_{i \in  N(g')}Y_{i}(g,u_{g'})&=&1.\nonumber\\
\Rightarrow  (n(g_{_1})-n(g')) Y_{k}(g,u_{g'})+ n(g')Y_{i}(g,u_{g'})  &=& 1.
\end{eqnarray}
\noindent Using induction hypothesis, we finally get  
\begin{equation}
Y_{k}(g,u_{g'})= \dfrac{1-\alpha}{n(g_{1})}, \;\;\textrm{for all $i \in N(g_{1}) \setminus N(g')$},
\end{equation}
 and  
\begin{equation}
Y_{i}(g,u_{g'}) =\dfrac{1-\alpha}{n(g_{1})}+\dfrac{\alpha}{n(g')},~\text{for all}~ i\in N(g').
\end{equation} 
Therefore, in either case, we have 
\begin{equation}
Y_{i}(g,u_{g'}) = {Y_{i}^{\alpha-CEM}(g,u_{g'})}.
\end{equation}
\textbf{Case~(b)}: $g' \supset g$.\\
Consider the pair $u_{g'}, \pi u_{g'} \in \mathbb H^N$. By Component Balance, Component-wise Weak Monotonicity and Anonymity, we have
\begin{eqnarray*}
\sum_{i \in N(g)}Y_{i}(g,u_{g'})=0 &\implies& N(g)Y_{i}(g,u_{g'})=0 \\
&\implies& Y_{i}(g,u_{g'})=0={Y^{\alpha -CEM}_{i}(g,u_{g'})}.
\end{eqnarray*}
\textbf{Case~(c)}: $g' \not\subseteq g$, but  $N(g') \cap N(g) \neq \emptyset$. Let $N(g') \cap N(g) =M$. \\
Now, for $i, j, \in N$, and $S \subseteq N(g)\setminus \{i,j\}$, we have
\begin{equation}
u_{g'}(g|_{S \cup i}) = u_{g'}(g|_{S \cup j}) = 0.
\end{equation}
In view of remark~\ref{rem:3}, it follows that 
\begin{equation}
Y_{i}(g,u_{g'})= Y_{j}(g,u_{g'}).
\end{equation}
By Component Balance and Anonymity, for each $g\in \mathbb G^N$, we have,
\begin{eqnarray*}
\sum_{i \in N(g)}Y_{i}(g,u_{g'})= u_{g'}(g)= 0 \implies n(g) Y_{i}(g,u_{g'})=0&& \\
\implies Y_{i}(g,u_{g'}) = {Y^{\alpha -CEM}_{i}(g,u_{g'})}\forall\;i \in N(g).&&
\end{eqnarray*}
This completes the proof.
\end{proof}  
\begin{remark}\rm \label{RM-2}
The four axioms of theorem \ref{thm3} are logically independent as shown below:
\begin{enumerate}[~(i)]
\item $Y(g,v)=0$ for all $g \in \mathbb G^N$ and $v\in \mathbb H^N$, satisfies all the axioms except Component Balance. 
\item  $Y(g,v)=Y_{i}^{MV}(g,v)$ if $v(g) \leq 5$ and {$Y(g,v)=Y_{i}^{CE}(g,v)$} if $v(g) > 5$ satisfies all the axioms except Linearity whenever $g \in \mathbb G^N$ and $v \in \mathbb H^N$..
\item $Y^{(-2)-MV}(g,v) =(-2)Y^{MV}(g,v) + 3 Y^{CE}(g,v) $ satisfies all the axioms except Component-wise Weak Monotonicity.
\item $Y(g,v)=\alpha_{i}v(h) $ such that $\alpha_{i} \neq \alpha_{j}$ for $i \neq j, ~~i,j \in N(h),~h \in C(g)$ , $\sum_{i \in N(h)} \alpha_{i}=1$ and $\alpha_{i} \geq 0$, satisfies all the axioms except Anonymity.
\end{enumerate}
\end{remark}
\noindent The following result ensures that Linearity can be replaced by Additivity in the characterization of the component-wise egalitarian Myerson value in the presence of Component-wise Local Monotonicity.
\begin{proposition}\rm \label{prp4}
Every Component Balanced allocation rule $Y:\mathbb G^N \times \mathbb H^N \mapsto \mathbb R^N$ satisfying  Additivity and Component-wise Local Monotonicity also satisfies Linearity.
\end{proposition}
\begin{proof}
Let a Component Balanced network allocation rule  $Y$ satisfy  Additivity and Component-wise Local Monotonicity. It is sufficient to show that $Y$ satisfies homogeneity, namely, 
\begin{equation}\label{eq:24}
Y(g,\lambda\cdot v)=\lambda\cdot Y(g,v)~~\forall \lambda \in \mathbb R, v \in \mathbb H^N.
\end{equation} 
First we show that Additivity and Component-wise Local Monotonicity imply Component-wise Strong Differential Monotonicity. \\
Let $v$ and $w$ be two component additive games and  {$i,j \in N(h)$, $h \in C(g)$} be such that
\begin{align*}\label{eq23}
v({g|_{S \cup i}})-v({g|_{S \cup j}}) \geq w({g|_{S \cup i}})-w({g|_{S \cup j}}), \tag{23} ~\forall {S \subseteq N(h) \setminus\{i,j\}}.
\end{align*}
It follows from Eq.(\ref{eq23}), that $(v-w)(g|_{S \cup i})\geq (v-w)(g|_{S \cup j}),  ~~{\forall S \subseteq N(h) \setminus\{i,j\}.}$ Thus, applying Additivity and Component-wise Local Monotonicity of $Y$ on the games $w$ and $v-w$, we get
\begin{align*}
Y_{i}(g,v)-Y_{j}(g,v)&=Y_{i}(g,w+(v-w))-Y_{j}(g,w+(v-w))\\
&=Y_{i}(g,w)-Y_{j}(g,w)+Y_{i}(g,v-w)-Y_{j}(g,v-w)\\
&\geq Y_{i}(g,w)-Y_{j}(g,w). 
\end{align*}
Thus, $Y$ satisfies Component-wise Strong Differential Monotonicity.\\
Now, we use a well known result on the linear space of real numbers that for a rational $\alpha \in \mathbb Q$, Additivity implies homogeneity. Thus, it remains to prove Eq.(\ref{eq:24}) only for $\lambda \in \mathbb{R \setminus Q}$.\\
Recall from Eq.(\ref{eq:1}) that every $v\in \mathbb H^N$ has a unique linear combination of the unanimity coefficients  $\Delta_{g'}$ and the unanimity games $u_{g'}$ such that $g'  \in \mathbb G^N$. The unanimity coefficients  $\Delta_{g'}$ are all zeroes whenever $g'$ contains more than one component. Thus, it is sufficient to prove Eq.(\ref{eq:24}) on $u_{g'}$ where $g'$ is connected and $\lambda \in \mathbb{R \setminus Q}$. Let, without loss of generality, $\lambda>0$. Also, because of Component Balance, there is no loss of generality in considering $g$ as connected. Since the rational numbers are dense in the reals, there are rational sequences {$(\lambda_{k}^{-})_{k \in \mathbb N}$ and $(\lambda_{k}^{+})_{k\in \mathbb N}$} such that $0<\lambda_{k}^{-}\leq \lambda \leq \lambda_{k}^{+} ~~{\forall k \in  \mathbb N}$, and $\displaystyle\lim_{k\rightarrow\infty}\lambda_{k}^{-}=\lim_{k\rightarrow\infty}\lambda_{k}^{+}=\lambda$.
\\For all $i \in N(g') \subseteq N(g)$, it can be easily shown that,
\begin{align*}
\lambda_{k}^{-}[u_{g'}(g|_{S \cup i})-u_{g'}(g|_{S\cup j})] &\leq\lambda [u_{g'}(g|_{S\cup i})-u_{g'}(g|_{S\cup j})] \\
&\leq \lambda_{k}^{+}[u_{g'}(g|_{S\cup i})-u_{g'}(g|_{S\cup j})] 
\end{align*}
for all  $S \subseteq N(g) \setminus\{i,j\}$ and $j \in N(g).$ 
\\Hence, $i,j$,  $\lambda_{k}^{-}.u_{g'}$ and $\lambda.u_{g'}$ satisfy the conditions of Component-wise Strong Differential Monotonicity. It follows that,
\begin{align*}
(Y_{i}(g,\lambda_{k}^{-}u_{g'})-Y_{j}(g,\lambda_{k}^{-}u_{g'}))&\leq (Y_{i}(g,\lambda .u_{g'})-Y_{j}(g,\lambda_.u_{g'}))\\
&\leq(Y_{i}(g,\lambda_{k}^{+}u_{g'})-Y_{j}(g,\lambda_{k}^{+}u_{g'})). 
\end{align*}
Since, Additivity implies homogeneity for rational scalars, we obtain
\begin{align*}
\lambda_{k}^{-}(Y_{i}(g,u_{g'})-Y_{j}(g,u_{g'}))&\leq (Y_{i}(g,\lambda .u_{g'})-Y_{j}(g,\lambda_.u_{g'}))\\
&\leq \lambda_{k}^{+}(Y_{i}(g,u_{g'})-Y_{j}(g,u_{g'})).
\end{align*}
Taking the limit and using our assumption, we have
\begin{align}
Y_{i}(g,\lambda.u_{g'})-Y_{j}(g,\lambda.u_{g'})=\lambda(Y_{i}(g,u_{g'})-Y_{j}(g,u_{g'})).
\end{align}
Summing up over $j \in N(g)$ we get,
\begin{align}
n(g)Y_{i}(g,\lambda.u_{g'})-\sum_{k \in N(g)}Y_{k}(g,\lambda.u_{g'})=n(g)\lambda.Y_{i}(g,u_{g'})-\lambda\sum_{k \in N(g)}Y_{k}(g,u_{g'}).
\end{align}
Note that $u_{g'}$ is component additive \footnote{See, for example, corollary 1 of lemma 1, page 268 in \cite{Nouweland}.}. Moreover, any  Component Balanced network allocation rule satisfies Efficiency. Therefore, using Component Balance, we get,
\begin{align*}
Y_{i}(g,\lambda.u_{g'})=\lambda.Y_{i}(g,u_{g'}).
\end{align*}
Similarly, we can show that Eq.(\ref{eq:24}) {holds  for} $i \in N(g) \setminus N(g')$  and for $\lambda<0.$
\end{proof}
\begin{remark}\rm
In view of proposition \ref{prp3} and proposition \ref{prp4}, the Linearity assumption in theorem~\ref{thm3} can be weakened by Additivity.  Thus, we have the second characterization theorem without proof as follows.
\end{remark}
\begin{theorem}\rm \label{thm:3(a)} 
A network allocation rule $Y:\mathbb G^N \times \mathbb H^N \mapsto \mathbb R^N $ is Component Balanced and satisfies Anonymity, Additivity, and Component-wise Weak Monotonicity if and only if it is the $\alpha$-component-wise egalitarian Myerson value $Y^{\alpha-CEM}$ for some $\alpha \in [0,1]$.
\end{theorem}
\noindent The next characterization requires Additivity, Component-wise Local Monotonicity, and the Superfluous Player in a Productive Network Property.
\begin{theorem}\label{them:4}\rm 
A Component Balanced network allocation rule $Y:\mathbb G^N \times \mathbb H^N \mapsto \mathbb R^N$ satisfies Additivity, Symmetry, Component-wise Local Monotonicity, and Superfluous Player in a Productive Network Property if and only if there exists an $\alpha \in [0,1]$ such that $Y(g,v)=Y^{\alpha-CEM}(g,v)$.\label{thm4} 
\end{theorem}
\begin{proof}
It is easy to show that $Y^{\alpha-CEM}$ is Component Balanced and satisfies Additivity, Component-wise Local Monotonicity and Superfluous Player in a Productive Network Property. For the uniqueness, assume that an allocation rule $Y$ satisfy these axioms.
\newline Let $v_{0}$ be the null game where for each $g \in \mathbb G^N$, every player is a superfluous player and, therefore, by Superfluous Player in a Productive Network Property, we have 
{$Y_i(g, v_0)\geq 0$} for all $i \in N$. Then following Component Balance and Component-wise Local Monotonicity, $ Y_{i}(g, v_{0})=0= {Y_{i}^{\alpha -CEM}(g, v_{0})}, ~~\forall i \in N(g)$.\\
Following similar arguments as in the proof of theorem~\ref{thm3}, it suffices to show the uniqueness of the $\alpha$-component-wise egalitarian Myerson value only on the unanimity games $u_{g'}$, where $g' \in \mathbb G^N$ is connected. Thus, for $g', g \in  \mathbb G^N$, as in theorem~\ref{thm3}, we have the following  cases: \\ 
\textbf{Case~(a)}: $g' \subseteq g$. First, take $g$ be connected. We have the following sub-cases: \\ 
\textit{Sub-case ($i$)}: $g' \subset g$. Observe that, $u_{g'}(g|_{S\cup i}) = u_{g'} (g|_{S\cup j})$ for all $i, j \in N(g')$ and $S \subseteq N \setminus \{i, j\}$.\\
Similarly, $u_{g'}(g|_{S \cup k}) = u_{g'}(g|_{S \cup p})$ for all $k, p \in N(g) \setminus N(g')$ and $S \subseteq N \setminus \{k, p\}$.\\
In view of remark~\ref{rem:3}, there exist $l, m \in \mathbb R$ such that
\begin{eqnarray}\label{Eq:04} 
Y_{i}(g,u_{g'}) =  Y_{j}(g,u_{g'}) = l,~~~~ \\
Y_{k}(g,u_{g'}) = Y_{p}(g,u_{g'}) = m.\footnote{If $n(g)-n(g')=1$, we simply take $Y_{k}(g,u_{g'})= m$ for $k \in N(g) \setminus N(g')$}
\end{eqnarray}
\noindent Also, we have,  $u_{g'}(g|_{S \cup i})\geq u_{g'}(g|_{S \cup k}) $ for $i \in N(g'), k \in N(g) \setminus N(g')$ and $S \subseteq N \setminus \{i, k\}$.
Therefore, again by Component-wise Local Monotonicity, we must have
\begin{equation}\label{Eq:02}
Y_{i}(g,u_{g'}) \geq Y_{k}(g,u_{g'}) \implies l \geq m.  
\end{equation}
\noindent  Moreover, $u_{g'}(g)= u_{g'}(g - g_k)$ for all $k \in N(g) \setminus N(g')$. Thus, by Superfluous Player in a Productive Network Property, 
\begin{equation}\label{Eq:01}
Y_{k}(g,u_{g'})= m \geq 0.
\end{equation} 
It follows that, $l \geq m \geq 0$. Let $l=s+m$ such that $s \geq 0$.
Applying Component Balance in Eq.(\ref{Eq:04}) we get, 
\begin{eqnarray}
n(g')l+(n(g)-n(g'))m = 1 
&\Rightarrow&  n(g')s+n(g)m = 1 ~~~~~~~~~~~~~~~~~~~~~~~~~~~~~~~~~~~~~~~~~~~~~~~~~~~~~~~~~~~~~~~~~~~\nonumber \\
&\Rightarrow& m = \dfrac{1-n(g')s}{n(g)}.~~~~~~~~~~~~~~~~~~~~~~~~~~~~~~~~~~~~~~~~~~~~~~~~~~~~~~~~~~~~~~~~~~~~~~~~ 
\end{eqnarray}
Since $m \geq 0$, we must have $s \leq \dfrac{1}{n(g')}$. Since $s \geq 0$, there is an $\alpha \in [0,1]$ such that $s = \dfrac{\alpha}{n(g')}$. Consequently, $m= \dfrac{1-\alpha}{n(g)}$. It follows that, 
\begin{equation*}
Y_{i}(g,u_{g'})=\left\{\begin{array}{lllll}
         &\dfrac{1-\alpha}{n(g)} + \dfrac{\alpha}{n(g')}\;\;\;\;\textrm{ for $i \in N(g')$}\\
         &~\\
         &\dfrac{1-\alpha}{n(g)} \;\;\;\;\textrm{for $i \in N(g) \setminus N(g')$}\\
         &~\\
         &0\;\;\;\;\textrm{for $i \notin N(g)$}.
         \end{array}\right.
\end{equation*}
Hence, $Y_{i}(g,u_{g'})= {Y_{i}^{\alpha -CEM}(g,u_{g'})}$ for all $i \in N(g)$ such that $g' \subset g $ and $Y_{i}(g,u_{g'})=0$ otherwise. This completes the proof of \textit{Sub-case (i)}.\\
\textit{Sub-case} ($ii$): Let $g'=g$. The result now follows from Component-wise Local Monotonicity and Component Balance. Component-wise Local Monotonicity and remark~\ref{rem:3} give 
\begin{equation}
Y_{i}(g,u_{g'}) = Y_{j}(g,u_{g'}),\;\;\textrm{for all $i,j \in N(g)$}.
\end{equation}
 Consequently, $Y_{i}(g,u_{g'})=l$, for some $l \in \mathbb R$ and all $i \in N(g)$. By Component Balance, we get 
\begin{equation*}
\displaystyle\sum_{i \in N(g)}Y_{i}(g,u_{g'})=l \cdot n(g)=1.
\end{equation*}
This would imply that $$Y_{i}(g,u_{g'})=\frac{1}{n(g)}= {Y_{i}^{\alpha -CEM}(g,u_{g'})},\;\;\textrm{for all $i \in N(g)$, and $\alpha \in [0,1]$.}$$
Next, consider $g' \subseteq g$ such that $g$ contains more than one component. Without loss of generality, we assume $g=g_{1} + g_{2}$ and $g_1 \cap g_2 =\emptyset$. Moreover, since $g'$ is connected, therefore, either we have $g' \subseteq g_{1}$ or $g' \subseteq g_{2}$.  {Assume that $g' \subseteq g_{1}$}. Recall that $u_{g'}$ is also component additive, therefore, we must have $u_{g'}(g_{1} + g_{2})=u_{g'}(g_{1} )+ u_{g'}( g_{2})=1$. Using Component Balance and Component-wise Local Monotonicity, we get,
\begin{eqnarray*}
\sum_{i \in N(g_{1}) \cup N(g_{2} )} Y_{i}(g,u_{g'})=1&\Rightarrow& \sum_{i \in N(g_{1} )} Y_{i}(g,u_{g'})+\sum_{i \in  N(g_{2} )} Y_{i}(g,u_{g'})=1\\
&\Rightarrow&  1 + n(g_{2}) Y_{i}(g,u_{g'}) = 1 \;\;\forall i \in N(g_2)\\
&\Rightarrow&  n(g_{2}) Y_{i}(g,u_{g'})=0\;\;\forall i \in N(g_2).
\end{eqnarray*} 
\noindent Hence, $ Y_{i}(g,u_{g'})=0= {Y_{i}^{\alpha -CEM}(g,u_{g'})}$ , for all $i \in N(g_{2})$ with $N(g') \cap N(g_{2})=\emptyset$. \\
Again, by Component Balance and Component-wise Local Monotonicity, we obtain
\begin{eqnarray*}
\sum_{i \in N(g_{1} )} Y_{i}(g,u_{g'})=1 &\Rightarrow & \sum_{k \in N(g_{1}) \setminus N(g')}Y_{i}(g,u_{g'})+ \sum_{i \in  N(g')}Y_{i}(g,u_{g'})=1.\\
&\Rightarrow &  (n(g_{1})-n(g')) Y_{k}(g,u_{g'})+ n(g')Y_{i}(g,u_{g'})  = 1.
\end{eqnarray*}
\noindent By Superfluous Player in a Productive Network Property, we finally obtain 
\begin{equation}
Y_{k}(g,u_{g'})= \dfrac{1-\alpha}{n(g_{1})}, ~~~\text{for all $i \in N(g_{1}) \setminus N(g')$}
\end{equation}   and  
$$Y_{i}(g,u_{g'}) =\dfrac{1-\alpha}{n(g_{1})}+\dfrac{\alpha}{n(g')},~\text{for all}~ i\in N(g').$$ 
It follows that in either case, $Y_{i}(g,u_{g'}) = {Y_{i}^{\alpha-CEM}(g,u_{g'})} $.\\
\textbf{Case (b)}: $g' \supset g$. By Component Balance and Component-wise Local Monotonicity, we have for each $g\in \mathbb G^N$,
\begin{eqnarray*}
\displaystyle\sum_{i \in N(g)}Y_{i}(g,u_{g'})=0 &\implies& n(g)Y_{i}(g,u_{g'})=0 \\
&\implies& Y_{i}(g,u_{g'})=0= {Y^{\alpha -CEM}_{i}(g,u_{g'})}.
\end{eqnarray*}
\textbf{Case (c)}: $g' \not\subseteq g$ but $N(g') \cap N(g) \neq \emptyset$. Let $N(g') \cap N(g) =M$. Then, 
by Component Balance and Component-wise Local Monotonicity, for each $g\in \mathbb G^N$, we get 
\begin{eqnarray*}
\displaystyle\sum_{i \in N(g)}Y_{i}(g,u_{g'})=0 &\implies& m Y_{i}(g,u_{g'})+(n(g)-m)Y_{k}(g,u_{g'}) =0 \\
&\implies& Y_{i}(g,u_{g'})=0= {Y^{\alpha -CEM}_{i}(g,u_{g'})},\;\;\textrm{for all $i \in N(g)$.}
\end{eqnarray*}
Following the same arguments as in Case (a), we obtain that $Y_{k}(g,u_{g'})=\beta^{*} \geq 0$ and $Y_{i}(g,u_{g'}) \geq  Y_{k}(g,u_{g'})$, for all $i \in M, k \in N(g) \setminus M$  with $\beta^{*} \in \mathbb R$. This completes the proof.   
\end{proof}
\begin{remark}\rm \label{RM-3}
The four axioms in theorem \ref{thm4} are logically independent as shown below:
\begin{enumerate}
\item  $Y_{i}(g,v)=0$, for all {$i \in N(g)$}, $g \in \mathbb G^N$, $v \in \mathbb H^N$ satisfies all the axioms except Component Balance. 
\item  $Y(g,v)=Y^{MV}(g,v)$ if $v(g) \leq 5$ and $Y(g,v)=Y^{CE}(g,v)$ if $v(g) > 5$ satisfies all axioms except Additivity.
\item Let $h \in C(g)$. Also, suppose $i\in N(h)$ be a superfluous player and $j \in N(h)$ be a non-superfluous player with respect to $v \in \mathbb H^N$. Take the vector $\alpha = (\alpha_1,\cdots, \alpha_n)\in \mathbb R^N$ such that $\alpha_{i}=\alpha_{j} <0 $ if $v(g|_{S \cup i}) \geq v(g|_{S \cup j}) $ and  $\alpha_{i} <0  < \alpha_{j}$ if $v(g|_{S \cup j}) \geq v(g|_{S \cup i})$ with the condition  $\sum_{i \in N(h)}\alpha_{i}=1$. Then $Y_{i}(g,v)=\alpha_{i}v(h)$ does not satisfy Superfluous Player in a Productive Network Property.
\item $Y^{(-2)-MV}(g,v)= (-2)Y^{MV}(g,v)+3Y^{CE}(g,v)$ does not satisfy Component-wise Local Monotonicity.  
\end{enumerate}
\end{remark}
\section{A Bidding Mechanism}\label{sec:5}
In this section, we study the non-cooperative foundation of the component-wise egalitarian Myerson value following the work of ~\cite{slikker07}. { This non-cooperative foundation gives us a strategic approach to characterize the value. In this approach, the payoff of a cooperative value arises as a result of players' equilibrium behavior through a bidding mechanism. The bidding mechanism identifies the factors due to which the agents prefer to deviate from marginalism to egalitarianism. }\cite{slikker07} proposes three bidding mechanisms that implement the Myerson value, the Position value, and the component-wise equal division rule under subgame perfect Nash equilibrium, respectively. Inspired by this work, we propose our bidding mechanism that implements the component-wise egalitarian Myerson value under subgame perfect Nash equilibrium. Our mechanism is also closely related to the implementation of the egalitarian Shapley value given by \cite{Brink}. Throughout the bidding mechanism, we assume that the network $g \in \mathbb G^N$ and the underlying network game $v \in \mathbb H^N$ are known to us. The mechanism is component-specific and recursive. Before describing the process, we present the following definitions and results due to \cite{slikker07}.
\begin{definition}~\cite{slikker07}
A network game $v \in \mathbb H^N$ is zero monotonic if  $v(g')-v(L_i(g'))\geq v(g'-L_i(g'))$ for $g' \subseteq g$ and $i \in N(g')$.
\end{definition} 
\begin{theorem}(\cite{slikker07}, Theorem 4.2, pp. 499)\label{thm:bid1}
Let $v$ be a component additive {network game} and $g$ a network. Then for all $ h \in C(g)$  , the payoff of player $i \in N(h)$ according to the Myerson value is given by, 
\begin{equation} Y_{i}^{MV}(g,v)=Y_{i}^{MV}(h,v)=\frac{1}{n(h)}\Big[v(h)-v(h-L_{i}(h))+\sum_{j \in N(h) \setminus i}  Y_{i}^{MV}(h-L_{j}(h),v)\Big]
\end{equation} 
\end{theorem} 
\begin{remark}\label{rm:EM}
Let $h_0\in C(g)$ such that $i \in N(h_0)$. Consider the value function $w : \mathbb H^N \mapsto \mathbb R$ given by $$w = \alpha v + (1-\alpha)\displaystyle\sum_{ h \in C(g)}v(h)\cdot u_{h}.$$ 
Then, using the fact that $Y^{MV}$ is Component Balanced and $Y_{i}^{MV}(g,u_h)=0$ ~{if}~ $i \notin N(h)$, due to Component Balance satisfied by $Y^{MV}$, we have 
\begin{eqnarray*}
Y_{i}^{MV}(g,w)&=& Y_{i}^{MV}\Big(g,\alpha v +(1-\alpha)\sum_{h \in C(g)} v(h)\cdot u_{h}\Big)\\
&=&\alpha Y_{i}^{MV}(g,v)+(1-\alpha)  Y_{i}^{MV}\Big(g,\sum_{h \in C(g)} v(h) \cdot u_{h}\Big)\\
&=&\alpha Y_{i}^{MV}(g,v)+(1-\alpha)  \Big[\sum_{h \in C(g)} v(h) Y_{i}^{MV}(g,u_h)\Big]\\
&=&\alpha Y_{i}^{MV}(g,v)+(1-\alpha) \frac{v(h_0)}{n(h_0)}~ 
\end{eqnarray*}
It follows that, $${Y_{i}^{\alpha - CEM}(g,v)}=Y_{i}^{MV}(g,w),\;\;\textrm{where $w= \alpha v + (1-\alpha)\sum_{h \in C(g)}v(h)\cdot u_{h}$}.$$
\end{remark}
\noindent In what follows next, we describe the bidding process for the component-wise egalitarian Myerson value.\\
\noindent\textbf{The bidding process:}\\
Given {$g = \emptyset$} (i.e., players in $g$ are isolated), all $i \in N(g)$  receive their  stand-alone payoff $0$. A non-zero value is generated only when the bidding takes place in a non-empty network. Assume that {$g \ne \emptyset$} and $n(h) > 1$ for each $h \in C(g)$ and let the mechanism has been specified for all components with at most $m-1$ {players}. We describe the rounds of the mechanism for a component $h \in C(g)$ with $m$ {players}. Depending upon the strategies of each player, the bidding process may have several rounds of bargaining, each consisting of four stages. Let $h_t$ be the component of $g$ with which the bidding process of round $t \in \{1,2,\cdots\}$ starts. After each round, the game ends, or new rounds start for all components remaining.
In the following, we describe these rounds in details.
{\textit{
\begin{enumerate} 
\item[\textbf{Round 1.}] $h_1=h$ ($t=1$). \textbf{Go to Stage 1}.
 \begin{enumerate}
\item[\textbf{Stage 1.}] Each player $i \in N(h)$ makes a bid $b_{j}^{i} \in \mathbb R$ for all $j \in N(h) \setminus i$. \\
Let $$B^{i}=\sum_{j \in N(h)\setminus i}(b_{j}^{i}-b_{i}^{j})$$ be the net bid of player $i$ measuring its `relative' willingness to be the proposer. \\
Let $i^{*}_{1}$ be the player with the highest net bid in this round. In case of a non-unique maximum value we choose any of these maximal bidders to be the `winner' with equal probability. \\
 $i^{*}_{1}$ pays every other player $j \in N(h) \setminus i^{*}_{1}$, the offered bid $b_{j}^{{i^{*}_{1}}}$. In the next stage $i^{*}_{1}$ being the ``winner" becomes the proposer.  \textbf{Go to Stage 2}.
\item[\textbf{Stage 2.}] Player $i^{*}_{1}$ proposes  offer $y_{j}^{i^{*}_{1}} \in \mathbb R$ to every player $j \in N(h) \setminus  i^{*}_{1}$. This offer is additional to the bid paid at {Stage 1}. \textbf{Go to Stage 3}.
\item[\textbf{Stage 3.}] Players other than  $i^{*}_{1}$ will sequentially accept or reject the  offer. If at least one player rejects it, the offer is rejected. Otherwise it is accepted. \textbf{Go to stage 4}.
\item[\textbf{Stage 4.}] The following two cases arise.
\begin{enumerate}
\item[\textbf{Case (a)}.] The offer is accepted. Then, each $j \in N(h) \setminus i^{*}_{1}$~ receives $y_{j}^{i^{*}_{1}}$, and player  $i^{*}_{1}$ obtains $$v(h)-\sum_{j \in N(h) \setminus i^{*}_{1}} y_{j}^{i^{*}_{1}}.$$ Hence  the final payoff  to player $j \in N(h) \setminus i^{*}_{1}$ is $$y_{j}^{i^{*}_{1}}+b_{j}^{i^{*}_{1}}.$$ The final payoff  to player $i^{*}_{1}$ is $$v(h)-\sum_{j \in N(h) \setminus i^{*}_{1}}(y_{j}^{i^{*}_{1}}+b_{j}^{i^{*}_{1}}).\;\;\;\;\;\textbf{ Stop}.$$ 
\item[\textbf{Case (b)}.] The offer is rejected. Then, player $i^{*}_{1}$ leaves the game with a probability $\alpha \in [0,1]$, and obtains her stand-alone payoff $v(L_{i^*_1}(h))$, while the players in $N(h)\setminus i^{*}_{1}$ enter into ~~~~~{Round 2} to bargain over $\alpha\cdot v(h- L_{i^*_1}(h))$. \\
Moreover, the game breaks down with a probability $(1-\alpha) \in [0,1]$ and all the players, including the proposer $i^{*}_{1}$, get zero payoff at this stage. Only the bids of {Stage 1} are transferred. It follows that the payoff to player $j \in N(h) \setminus i^{*}_{1}$ is $b_{j}^{i^{*}_{1}}$ and the payoff to proposer $i^{*}_{1}$ is $-\displaystyle\sum_{j \in N(h) \setminus i^{*}_{1}}b_{j}^{i^{*}_{1}}.$~~~~~~\textbf{Stop}.
\end{enumerate}
\end{enumerate}
\end{enumerate}
Note that the possibility of breaking down is a common prior to all the players. Moreover, if $h - L_{i^{*}_{1}}(h)$ is a network with multiple components rather than a single component then in the next round we have  multiple identical mechanisms in parallel for each such component.  \\
\noindent In the following rounds,  Stage 1,2 and 3 are identical to round 1 but  the player set is reduced as  the proposers in the previous rounds have left the game. However, unlike the first round there is no possibility of breakdown at the Stage 4 of the following rounds. To make this section self-contained we describe the $t$-th round as follows.
\begin{enumerate}
 \item[\textbf{Round t.}] $t \in \{2,\ldots,n-1\}: h_t=h_{t-1}- L_{i^{*}_{t-1}}(h)$. Suppose, without loss of generality, $h_t \in C(g)$. \\ \textbf{Go to Stage 1}.
\begin{enumerate}
\item[\textbf{Stage 1.}] Each player $i_{t} \in N(h_{t})$ offers bid $b_{j}^{i_{t}} \in \mathbb R$ for every {$j \in N(h_{t}) \setminus i_{t}$}. For  $i_{t} \in N(h_{t})$, let 
$$B^{i_t}=\sum_{j \in N(h) \setminus i_{t}} (b^{i_{t}}_{j}-b^{j}_{i_{t}}),$$
be the net bid of player $i_{t}$. Let $i^{*}_{t}$ be the player with the highest net bid (in Round $t$). In case of a non-unique maximum value of net bid we choose any of these maximal bidders to be the `winner' with equal probability. \\
Then player $i^{*}_{t}$ being the winner pays every other player $j \in N(h_{t}) \backslash i^{*}_{t}$ its offered bid $b_{j}^{{i^{*}_{t}}}$. In the next stage, $i^{*}_{t}$ becomes the proposer. \textbf{Go to Stage 2}.
\item[\textbf{Stage 2.}] Player $i^{*}_{t}$ proposes an offer $y_{j}^{i^{*}_{t}} \in \mathbb R$ to every player $j \in N(h_{t})\setminus i^{*}_{t}$. This offer is additional to the bid paid at Stage 1. \textbf{Go to Stage 3}.
\item[\textbf{Stage 3.}] The players except $i^{*}_{t}$ sequentially accept or reject the  offer. If at least one player rejects it, then the offer is rejected. Otherwise the offer is accepted. \textbf{Go to Stage 4}.
\item[\textbf{Stage 4.}] The following two cases arise.
\begin{enumerate}
\item[\textbf{Case (a).}] If the offer is accepted, then $j \in N(h_{t})\setminus i^{*}_{t}$ receives $y_{j}^{i^{*}_{t}}$ and  $i^{*}_{t}$ obtains $$v(h_t)-\sum_{j\in N(h_{t}) \setminus i^{*}_{t}} y_{j}^{i^{*}_{t}}.$$ Hence, in this case the final payoff to player $j \in  N(h_{t})\setminus i^{*}_{t}$ is $$y_{j}^{i^{*}_{t}}+b_{j}^{i^{*}_{t}}+\displaystyle\sum_{k=1}^{t-1}b_{j}^{i^{*}_{k}}.$$ 
On the other hand, player $i^{*}_{t}$ receives $v(h_t)-\sum_{j \in N(h_t) \setminus  i^{*}_{t}}(y_{j}^{i^{*}_{t}}+b_{j}^{i^{*}_{t}}).$ \textbf{ Stop}.\\
\item[\textbf{Case (b).}] If the offer is rejected then  player $i^{*}_{t}$ leaves the game and obtains  her stand-alone payoff, $v(L_{i^{*}_{t}}(h_t))$, while the players in $N(h_t)\setminus i^{*}_{t}$ proceed to round $t+1$ to bargain over $v(h_t -L_{i^{*}_{t}}(h_t))$. \textbf{ Stop}.
\end{enumerate}
\end{enumerate}
\end{enumerate}
}}
\noindent Note that the proposed mechanism allows the breakdown of the network during the bidding with some endogenous probability  $\alpha$ if the proposal is rejected in the first round. Given $v \in \mathbb H^N$, the final payoffs of the players who are assumed to be risk-neutral in the mechanism are calculated as follows. \\
When rejected in the first round, the expected final gain of proposer $i^*_1$ is given by $$\alpha \cdot \Big(v(L_{i^*_1}(h))- \displaystyle\sum_{j \ne i^*_1} b_j^{i^*_1}\Big).$$ 
On the other hand, every other player $j \in N(h) \setminus i^{*}_{1}$ finally obtains $b^{i^*_1}_j$ plus the expected payoff due to the contingent (with probability $\alpha$) outcome of the mechanism continuing with player set $N(h)\setminus i^*_1$, see \cite{Brink}. In case of acceptance of the proposal in the first round, the final gain of $i^*_1$ is $$v(h)- \displaystyle \sum_{j \in N(h) \setminus i^{*}_{1}}\Big(b^{i^*_1}_j + y^{i^*_1}_j\Big).$$ Consequently, the final gain of every other player $j$ is $b^{i^*_1}_j + y^{i^*_1}_j$.\\
In the following, we prove that, for any zero-monotonic network game such that the network generates a nonnegative worth, the given bidding mechanism implements the $\alpha$-component-wise egalitarian Myerson values as subgame perfect equilibrium (SPE) outcomes.
\begin{theorem}
Let {$g\in \mathbb G^N$}, $h\in C(g)$, and $\alpha \in [0,1]$ be the probability that the bidding continues after rejection in the first round. Also let $v \in \mathbb H^N$ be a {zero monotonic  network game}. Then the outcome in any sub-game perfect equilibrium  of the bidding mechanism coincides with the payoff vector {$Y^{\alpha - CEM}(h,v)$}.
\end{theorem}
\begin{proof}
Let $g \in \mathbb G^N$, $h\in C(g)$, and $v$ be a zero monotonic network  game, and let  $\alpha \in [0,1]$. First, we prove that the bidding mechanism implements {$Y^{\alpha - CEM}(h,v)$} for the zero monotonic network game $v$ in subgame perfect Nash equilibrium, and then we prove that any subgame perfect Nash equilibrium induces the component-wise egalitarian Myerson value.\\
The proof proceeds by induction on the number $k$ of players in the component. \\
The theorem holds for $k = 1$ trivially.\\
Let the theorem hold for $k = m-1$. We show that it also holds for $k= m$. \\
We first prove that the outcome prescribed by the component-wise egalitarian Myerson value is a subgame perfect Nash equilibrium outcome. We explicitly construct a subgame perfect Nash equilibrium that yields this value as a subgame perfect Nash equilibrium outcome. We describe the stages of each round as follows.
\begin{enumerate}
\item[\textbf{Round~1.}]~~ $h_t = h$.
\begin{enumerate} 
\item[\textbf{Stage 1.}] For every $j \in N(h) \setminus  i$, each player $i \in N(h)$, announces 
\begin{equation}
b_{j}^{i}=\alpha (Y^{ MV}_{j}(h,v)-Y^{ MV}_{j}(h-L_i(h),v))+\frac{v(h)}{n(h)}(1-\alpha)
\end{equation} 
Let $i^{*}_{1}$ be a player with highest net bid. Once proposer $i^{*}_{1}$ is determined, she pays every other player $j \in N(h) \setminus i_{1}^{*}$ her offered bidding amount, namely\footnote{This bid is decided on the basis of remark~\ref{rm:EM}.}
\begin{equation} b_{j}^{i^{*}_{1}}=\alpha (Y^{ MV}_{j}(h,v)-Y^{ MV}_{j}(h-L_i(h),v))+\frac{v(h)}{n(h)}(1-\alpha)
\end{equation} 
\item[\textbf{Stage 2.}] Along with the bid, the proposer offers $y_{j}^{i^{*}_{1}}=\alpha Y^{MV}_{j}(h-L_{ i^{*}_{1}}(h),v)$ to every $j \in N(h) \setminus  i^{*}_{1}$.
\item[\textbf{Stage 3.}] Each $j \in N(h) \setminus  i^{*}_{1}$ accepts the offer if $y_{j}^{i^{*}_{1}} \geq \alpha Y^{MV}_{j}(h-L_{ i^{*}_{1}}(h),v)$ and rejects otherwise.\\
\noindent In case of acceptance, all $j \in N(h)$ get payoff $Y^{\alpha-CEM}_{j}(h,v)$ and in case of rejection with probability $\alpha$ player $i^{*}_{1}$ leaves the game with a payoff $\alpha v(L_{i^{*}_{1}}(h))$ and the remaining players play the second round with a total worth of $\alpha v(h - L_{ i^{*}_{1}}(h))$. 
\end{enumerate} 
\item[\textbf{Round t.}] Let $t \in \{2,3,\ldots,m-1\}$. Take $N(h_{t})= N(h_{t-1})\setminus i^{*}_{t-1}$. \textbf{ Go to stage 1}.
\begin{enumerate}
\item[\textbf{Stage 1.}] For every $j \in N(h_{t})\setminus i_{t}$, player $i_{t} \in N(h_{t})$, announces the following bid 
\begin{equation} b_{j}^{i_{t}}=\alpha (Y^{MV}_{j}(h_{t},v)-Y^{MV}_{j}(h_{t}-L_{i_{t}}(h_t),v))
\end{equation}  
 Suppose $i^{*}_{t}$ be a player with maximum net bid and is selected as the winner. Then $i^{*}_{t}$ pays every  $j \in N(h_{t})\setminus i_{t}$ the offered bidding amount, namely 
\begin{equation}
b_{j}^{i^{*}_{t}}=\alpha (Y^{MV}_{j}(h_{t},v)-Y^{MV}_{j}(h_t-L_{i^{*}_{t}}(h_t),v)).
\end{equation}
\item[\textbf{Stage 2.}] Along with the bid, the proposer offers to every player $j \in N(h) \setminus  i^{*}_{t}$ the following amount: 
\begin{equation}
y_{j}^{i^{*}_{t}}=\alpha Y^{MV}_{j}(h_{t}- L_{i^{*}_{t}}(h_t),v).
\end{equation}
\item[\textbf{Stage 3.}] The players $j  \in N(h_{t} - L_{i_t}(h_t))$ accept the offer whenever $y_{j}^{i^{*}_{t}} \geq \alpha Y^{MV}_{j}(h_{t}-L_{ i^{*}_{t}}(h),v)$ and reject otherwise. 
In case of acceptance, the final payoff of $j \in  N(h_{t}) \setminus i_{t}$ is 
\begin{equation} y_{j}^{i^{*}_{t}}+b_{j}^{i^{*}_{t}}+ \sum_{k=1}^{t-1}b_{j}^{i^{*}_{k}}={Y_{j}^{\alpha -CEM}(h,v)}
\end{equation}
\noindent  and player $i^{*}_{t}$ receives 
\begin{equation}
v(h_{t})- \sum_{j \in  N(h_{t}) \setminus i_{t}}(y_{j}^{i^{*}_{t}}+b_{j}^{i^{*}_{t}}+ \sum_{k=1}^{t-1}b_{j}^{i^{*}_{k}})={Y_{i^{*}_{t}}^{\alpha -CEM}(h_t,v)}
\end{equation} 
In case of rejection player $i^{*}_{t}$ leaves the game with $\alpha v(L_{i^{*}_{t}}(h))$ and the remaining  player play the next round with total worth $\alpha v(h_{t}-L_{i^{*}_{t}}(h))$.

\noindent Since the Myerson value satisfies Balanced Contribution, therefore the net bid becomes, 
\begin{eqnarray*}
B^{i}&=&  \sum_{j \in N(h) \setminus  i}(b_{j}^{i}-b_{i}^{j})\\
&=&  \sum_{j \in N(h) \setminus  i}\alpha (Y_{j}^{MV}(h,v)-Y_{j}^{MV}(h-L_{i}(h),v))+\frac{v(h)}{n(h)}(1-\alpha)\\
&&- \sum_{j \in N(h) \setminus  i}\alpha(Y_{i}^{MV}(h,v)-Y_{i}^{MV}(h-L_{i}(h),v))-\frac{v(h)}{n(h)}(1-\alpha)\\
&=& \alpha\Bigg[ \sum_{j \in N(h) \setminus  i}(Y_{j}^{MV}(h,v)-Y_{j}^{MV}(h-L_{i}(h),v))-(Y_{i}^{MV}(h,v)-Y_{i}^{MV}(h-L_{ j}(h),v))\Bigg]\\
&=& 0.
\end{eqnarray*} 
\end{enumerate}
\end{enumerate}
\noindent Next we make the following claims,
\begin{enumerate}
\item[\textbf{Claim 1.}] The strategies constitute a sub-game perfect Nash equilibrium. \\
The given game being zero monotonic, we prove the following assertions.
\begin{enumerate}
\item[\textbf{(i)}] At Stage 3, it is optimal for the player $j \in N(h)\setminus i^{*}_{1}$ to accept $y_{j}^{i^{*}_{1}}$ if 
\begin{equation}
 y_{j}^{i^{*}_{1}}\geq   \alpha Y_{j}^{MV}(h- L_{i^{*}_{1}}(h),v)
\end{equation}
and reject otherwise. If $j$ accepts the offer $y_{j}^{i^{*}_{1}}$ such that $$y_{j}^{i^{*}_{1}} < \alpha Y_{j}^{MV}(h- L_{i^{*}_{1}}(h),v)$$  in  Round 1, then she ends up  with a total payoff less than {$ Y_{j}^{\alpha -CEM}(h,v)$}. If she rejects, in the next round she would definitely get {$Y_{j}^{\alpha -CEM}(h,v)$} in total. 
\item[\textbf{(ii)}]  At Stage 2, it is optimal for the player $i^{*}_{1}$ to offer 
\begin{equation} 
y_{j}^{i^{*}_{1}} = \alpha Y_{j}^{MV}(h- L_{i^{*}_{1}}(h),v).
\end{equation}
\noindent  If she offers $y_{j}^{i^{*}_{1}} < \alpha Y_{j}^{MV}(h- L_{i^{*}_{1}}(h),v)$, any other player $j \in N(h)\setminus i^{*}_{1}$ rejects the offer immediately due to \textbf{(i)}.\\
If $y_{j}^{i^{*}_{1}} > \alpha Y_{j}^{MV}(h- L_{i^{*}_{1}}(h),v)$ then all $j \in N(h) \setminus i^{*}_{1}$ accept the offer immediately as their final payoff would be greater than $\alpha Y_{j}^{MV}(h,v)$ but player $i^{*}_{1}$ ends up with final payoff less than $\alpha Y_{i^{*}_{1}}^{MV}(h,v)$. 
Therefore, she has no incentive to be a proposer, as in this case, she benefits more by not being a proposer.
\item[\textbf{(iii)}] The strategies of Stage 1 constitute a sub-game perfect Nash equilibrium.\\
\noindent  Note that if any one of the players increases her bid $\displaystyle\sum_{j \in N(h)\setminus i^{*}_{1}}b^{i^{*}_{1}}_{j}$, she will be chosen as the proposer for sure, but her payoff will decrease and if she decreases her total bid then any one of the remaining players becomes a proposer so that she finally obtains the same payoff {$ Y_{i}^{\alpha -CEM}(h,v)$} in the end; hence, she remains indifferent in changing the bid.
\noindent  This proves our assertion.
\end{enumerate}
\item[\textbf{Claim 2.}] Any sub-game perfect Nash equilibrium yields the $\alpha$-CEM value. 
Consider the case with $n(h)=m > 1$. By method of induction, assume that the mechanism implements ${Y^{\alpha -CEM}(h - L_{i^{*}_{1}}(h),v )}$ for the sub-network $h- L_{i^{*}_{1}}(h)$. We prove the following assertions.
\begin{enumerate}
\item[\textbf{(i)}]  At stage 3, all players except the proposer $i^{*}_{1}$ accept the offer $y_{j}^{i^{*}_{1}}$ such that 
\begin{equation}
 y_{j}^{i^{*}_{1}}\geq \alpha Y^{MV}_{j}(h- L_{i^{*}_{1}}(h),v).
\end{equation}
If~ $y_{j}^{i^{*}_{1}} < \alpha Y^{MV}_{j}(h-L_{i^{*}_{1}}(h),v)$,~ the offer is  rejected as by the induction hypothesis in the next round, $j$ would get $ Y^{MV}_{j}(h- L_{i^{*}_{1}}(h),v)$ after rejection. \\
Let $\beta $ be the last player who would decide whether to accept or  {reject} the offer, if $$y_{\beta}^{i^{*}_{1}} < \alpha Y^{MV}_{\beta}(h- L_{i^{*}_{1}}(h),v),$$ 
then $\beta$ will reject the offer. The second last player $\beta -1$ anticipates the reaction of $\beta$, if $$y_{\beta -1}^{i^{*}_{1}} > Y^{MV}_{\beta -1}(h- L_{i^{*}_{1}}(h),v)\;\;\textrm{and}\;\; y_{\beta}^{i^{*}_{1}} > Y^{MV}_{\beta}(h- L_{i^{*}_{1}}(h),v)$$ and when the game reaches $\beta -1$, she will accept the offer. If $$y_{\beta -1}^{i^{*}_{1}} < \alpha Y^{MV}_{\beta -1}(h- L_{i^{*}_{1}}(h),v)\;\;\textrm{and}\;\; y_{\beta}^{i^{*}_{1}} > \alpha Y^{MV}_{\beta}(h- L_{i^{*}_{1}}(h),v),$$ she will reject the offer and if 
$$y_{\beta}^{i^{*}_{1}} < \alpha Y^{MV}_{\beta}(h- L_{i^{*}_{1}}(h),v), $$ she is indifferent to acceptance or rejection. This is because, she knows that $\beta$ is any how going to reject the offer.
\item[\textbf{(ii)}]  If $v(h) > v(h - L_{i^{*}_{1}}(h))+v(L_{i^{*}_{1}}(h))$ then at Stage 2 the proposer will offer $$y_{j}^{i^{*}_{1}} = Y^{MV}_{j}(h- L_{i^{*}_{1}}(h),v).$$
Following  induction hypothesis, proposer $i^{*}_{1}$ knows  that the players will not reject this offer. Note that, unlike the Myerson value, here if $$v(h) = v(h - L_{i^{*}_{1}}(h))+ v(L_{i^{*}_{1}}(h)),$$ then rejection will not constitute a subgame perfect Nash equilibrium. Because, after rejection, player $ i^{*}_{1}$ will get $\alpha v(L_{i^{*}_{1}}(h))\leq v( L_{i^{*}_{1}}(h))$. Hence, for $$v(h) \geq v(h - L_{ i^{*}_{1}}(h))+v(L_{i^{*}_{1}}(h)),$$ the rejection of the offer made by $i^{*}_{1}$ cannot be a subgame perfect Nash equilibrium. \newline 
 In case of rejection, the expected payoff of player $i^{*}_{1}$  is $\alpha v(L_{i^{*}_{1}}(h))$. She can improve her payoff by offering 
$$\alpha Y^{MV}_{j}(h- L_{i^{*}_{1}}(h),v)+\frac{\epsilon}{n(h)-1}$$ to all $j \in N(h) \setminus i^{*}_{1}$ with $0< \epsilon < v(h) - v( h - L_{i^{*}_{1}}(h))-v(L_{i^{*}_{1}}(h))$ so that her offer is accepted (by Claim (1)). Therefore, a subgame perfect Nash equilibrium requires acceptance of the proposal. This implies that, for all $j \in N(h) \setminus i^{*}_{1}$, we must have 
$$y_{j}^{i^{*}_{1}} \geq \alpha Y^{MV}_{j}(h-L_{i^{*}_{1}}(h),v).$$ 
However, the offer $y_{j}^{i^{*}_{1}} > \alpha Y^{MV}_{j}(h- L_{i^{*}_{1}}(h),v)$ cannot be a part of a subgame perfect Nash equilibrium, since $i^{*}_{1}$ could still offer $\alpha  Y^{MV}_{j}(h- L_{i^{*}_{1}}(h),v)+\frac{\epsilon}{n-1} $ to all $j \neq i^{*}_{1}$ with $0< \epsilon < y_{j}^{i^{*}_{1}} - \alpha Y^{MV}_{j}(h- L_{i^{*}_{1}}(h),v) $. These offers are accepted and the payoff to $i^{*}_{1}$ increases. Therefore,  $y_{j}^{i^{*}_{1}} = \alpha Y^{MV}_{j}(h- L_{i^{*}_{1}}(h),v)$ , for all $j \neq i^{*}_{1}$ at any subgame perfect Nash equilibrium. Finally acceptance of the proposal implies that at Stage $3$, every agent $j \in N(h) \setminus i^{*}_{1}$ accepts an offer $y_{j}^{i^{*}_{1}}$ such that $$ y_{j}^{i^{*}_{1}}\geq \alpha Y^{MV}_{j}(h- L_{i^{*}_{1}}(h),v).$$
\item[\textbf{(iii)}]  In any subgame perfect Nash equilibrium, $B^{i}=B^{j}$, for all $i$ and $j$ and, therefore, $B^{i}=0$, for all {$i \in N(h)$}.\\
\noindent Denote by $\Omega=\{i \in N(h)|B^{i}=\max_{j}B^{j}\}$, if $\Omega = N(h)$, the claim holds true,  {since $\sum_{i \in N(h)}B^{i}=0$}. Otherwise, we can establish that any player $i \in \Omega$ can change her bid so as to decrease the sum of payments in case she wins. Furthermore, these changes can be made without changing the set $\Omega$. Let $\sum_{i \in N(h)}B^{i}\neq 0$.
Any player $i \in \Omega$ can change her bid to decrease the sum of payment if she wins. Moreover, these changes can be done without altering the set $\Omega$. Therefore, the proposer can maintain the same probability of winning and obtain a higher expected payoff. Formally the process is as follows. \\
Let $j \notin \Omega$, and $i \in \Omega$ change her strategy by announcing 
$$b_{k}^{i'}=b_{k}^{i}+\delta,\;\;\textrm{for all $k \in \Omega$ and $k \neq i$,}$$ 
$$b_{j}^{i'}=b_{j}^{i}-|\Omega|\delta \;\;\textrm{and}\;\; b_{l}^{i'}=b_{l}^{i},\;\;\textrm{for all $l \notin \Omega$ and $l \neq j$.}$$ 
The new net bids are $$B^{i'}=B^{i}-\delta\;\;\textrm{and $B^{k'}=B^{k}-\delta,~ \forall ~ k \in \Omega$ and $k \neq i$,}$$ 
$$B^{j'}=B^{j}+|\Omega| \delta\;\;\textrm{and $B^{l'}=B^{l}$ for $l \notin \Omega$ and $l \neq j$.}$$ 
If $\delta $ is small enough such  that $B^{j}+|\Omega| \delta < B^{i}- \delta$, \footnote{Recall that $B^{j}<B^{i}$} then 
$$B^{l'}<B^{i'}=B^{k'}\;\;\textrm{for all $l \notin \Omega$ (including $j$) and for all $k \in \Omega$.}$$ Hence $\Omega$ does not change. {However, $\sum_{p \neq i}b_{p}^{i}-\delta < \sum_{p \neq i}b_{p}^{i}$}.
\item[\textbf{(iv)}]  In any subgame perfect Nash equilibrium, each player's payoff is the same regardless of who is chosen as proposer. \\
\noindent  We already know that all bids $B^{i}$ are the same. If player $i$ would strictly prefer to be the proposer, she could improve her payoff by slightly increasing one of her bids $b_{j}^{i}$. If $i$ strictly wants that some other player $j$ should be the proposer, she can increase her payoff by decreasing $b_{j}^{i}$. Since $i$ does not do so in equilibrium, it follows that she is indifferent to the proposer's identity.
\item[\textbf{(v)}]  In any subgame perfect Nash equilibrium, the final payoff received by each of the players is her  {component-wise} egalitarian Myerson payoff.\\
\noindent Let $i$ be the proposer, her final payoff is given by $$x_{i}^{i}=v(h)-\alpha v(h - L_{i}(h))-\sum_{j \in N(h) \setminus i}b_{j}^{i}.$$ 
If $i$ be a player other  then the proposer $j$, her final payoff is $$x_{i}^{j}= \alpha Y_{i}^{MV}(h - L_{j}(h),v)+b_{i}^{j}.$$ Sum of the payoffs of player $i$ over all possible choices of proposers is given by,
\begin{eqnarray*}
\sum_{j \in N(h)} x_{i}^{j} & = & (v(h)-\alpha v(h -L_{ i}(h))-\sum_{j \in 
N(h) \setminus i} b_{j}^{i})+\sum_{j \in N(h) \setminus i}(\alpha Y^{MV}_{j}(h -L_{i}(h),v) +b^{j}_{i})\\
&=&  v(h)-\alpha v(h - L_{i}(h))+ \sum_{j \in N(h) \setminus i} \alpha Y^{MV}_{j}(h - L_{ i}(h),v)-B^{i}\\
&=& (1- \alpha)v(h)+\alpha(v(h)- v(h - L_{i}(h)))+\sum_{j \in N(h) \setminus i}\alpha Y^{MV}_{j}(h -L_{ i}(h)),v)\\
&=& (1- \alpha)v(h)+\alpha[(v(h)- v(h -L_{i}(h))+\sum_{j \in N(h) \setminus i} Y^{MV}_{j}((h - L_{ i}(h),v)]\\
&=& (1- \alpha)\frac{n(h)v(h)}{n(h)}+\alpha[n(h) Y_{i}^{MV}(h,v)]~(\text{by theorem}~\ref{thm:bid1})\\
&=& n(h) {Y^{\alpha-CEM}_{i}(h,v)}
\end{eqnarray*}
Therefore the payoff of player $i$ is {$Y^{\alpha- CEM}_{i}(h,v)$}. This completes the proof.
\end{enumerate}
\end{enumerate}
\end{proof} 
\begin{remark} 
Note that for $\alpha =0$ and $\alpha =1$, the bidding mechanism described above coincides with the bidding mechanism that implements $Y^{CE}(g,v)$ and   $Y^{MV}(g,v)$ respectively.\\
If the players have prior knowledge that there is no chance of breakdown at all, then they carry the bargaining procedure further, and at equilibrium, the players generate Myerson payoff, which is purely marginalistic. On the other extreme, whenever the players have prior knowledge that if at least one of them disagrees with the offer, then the game will break down. In this situation at equilibrium, the players generate component-wise egalitarian value. This is desirable because even though different players have different levels of productivity, but the presence of each player is indispensable for the successful completion of the bidding process. Hence they generate equal payoff at sub-game perfect Nash equilibrium. The probability $(1-\alpha)$ with which the bidding procedure is about to break in case of disagreement decides the degree of compassion exhibited by the players at equilibrium behavior.
\end{remark}
\section{Conclusions}\label{sec:6}
In this paper, we have first provided three characterizations of the component-wise egalitarian Myerson value for network games. The formulation is in line with the egalitarian Shapley value characterized by \cite{Casajus,Brink}.  We have also proposed a bidding mechanism for the class of component-wise egalitarian Myerson values following a similar procedure as that of \cite{Slikker05a}. { The proposed mechanism provides a strategic approach to the component-wise egalitarian Myerson value. Each stage of the bidding mechanism analyzes the players' strategic concerns under a cooperative environment. In this approach, the payoff of a cooperative value arises as a result of players' equilibrium behavior through the bidding mechanism. The bidding mechanism identifies the factors due to which the agents prefer to deviate from marginalism to egalitarianism, which is the main consideration of our present paper. The component-wise egalitarian Myerson value is a player-based allocation rule, meaning: they depend more on the players in the network, {see \cite{Jackson2005}}. Similar allocation rules that depend more on the links in a network, we call them link-based allocation rules are equally important, if not more, to investigate. Moreover, we see a possibility of extending this idea to hypergraph games which will be itself an interesting topic to study. These we keep for our future study.}
\subsection*{Acknowledgement}
The comments and suggestions of the AE and the two anonymous reviewers are highly appreciated. We also acknowledge the financial support from ASTEC Grant No. ASTEC/S\&T/192(171)/2019-20/2762.

\end{document}